# Scaling notifications beyond alerts: from subtly drawing attention up to forcing the user to take action

DENYS J.C. MATTHIES, Augmented Human Lab, University of Auckland, NZ. Fraunhofer IGD Rostock, GER
LAURA MILENA DAZA PARRA, Fraunhofer IGD Rostock, GER
BODO URBAN, Fraunhofer IGD Rostock, GER

New computational devices, in particular wearable devices, offer the unique property of always being available and thus to be able to constantly update the user with information, such as by notifications. While research has been done in sophisticated notifications, devices today mainly stick to a binary level of information, while they are either attention drawing or silent. In this paper, we want to go further and propose scalable notifications, which adjust the intensity reaching from subtle to obtrusive and even going beyond that level, while forcing the user to take action. To illustrate the technical feasibility and validity of this concept, we developed three prototypes providing mechano-pressure, thermal, and electrical feedback and evaluated them in different lab studies. Our first prototype provides subtle poking through to high and frequent pressure on the user's spine, which creates a significantly improved back posture. In a second scenario, the users are enabled to perceive the overuse of a drill by an increased temperature on the palm of a hand until the heat is intolerable and the users are forced to eventually put down the tool. The last project comprises a speed control in a driving simulation, while electric muscle stimulation on the users' legs conveys information on changing the car's speed by a perceived tingling until the system independently forces the foot to move. Although our selected scenarios are long way from being realistic, we see these lab studies as a means to validate our proof-of-concept. In conclusion, all studies' findings support the feasibility of our concept of a scalable notification system, including the system of forced intervention. While we envisage the implementation of our proof-of-concept into future wearables, more realistic application scenarios are worthy of exploration.

**Keywords:** Scaling notifications; electrical muscle stimulation; thermal feedback; haptic feedback; wearable prototyping; assistive technology.

## 1 INTRODUCTION

An essential property of ubiquitous computing (Weiser, 1997 [62]) is the omnipresence of computers that occur in all kinds of forms, such as wearable computers (Mann, 2001 [35]). The high availability of wearables resting on the human body all day is unique and allows constant conveying of information. Presently, using notifications to keep the user up-to date is an integral part of many wearables, such as smartwatches notifying the user about incoming messages, calls, activity goals, calendar events, alarms, etc. (Sahami Shirazi et al., 2014 [50]). Many new types of future wearables, for instance smart insoles, may substantially expand the range of features beneficial for users with special needs, such as elderly people or diabetics, while notifying them when entering dangerous areas (Matthies et al., 2017 [36]), or when dangerous foot postures occur (El-Sayed et al., 2011 [18]). While the sensing part is already technically feasible, we nevertheless need to ask the questions:

*How can we convey such information appropriately to the context to the user and how can we possibly intervene in an interaction scenario?*

In this paper, we outline possible answers by introducing the concept of *Scaling notifications beyond alerts* – a scalable notification system that, on the one hand, provides very subtle feedback to the user while, on the other hand, can intervene and force the user to take action.

While subtle notifications may be quickly overlooked or intentionally ignored without great effort and disruption, these types of notification are suitable for less important information. However, there may be very urgent notifications, such as important information, to keep ourselves healthy. These notifications should be a level beyond being obtrusive, in a way that they cannot be overlooked and where one is possibly forced to act.

Authors' addresses: Augmented Human Lab, Auckland Bioengineering Institute, The University of Auckland, 70 Symonds Street, Auckland, 1010, New Zealand. Fraunhofer IGD Rostock, Joachim-Jungius-Str. 11, 18059 Rostock, Germany. denys@ahlab.org,

## 2 RELATED WORK

Before introducing our concept and evaluation, we present a broad literature review; by introducing a general overview (2.1) on our senses used for the perception of notifications and the variations of forms and nuances of notifications (2.2), while illustrating these with exemplary previous works.

### 2.1 Background: Overview of human senses

As we can see from *Figure 1,* the human body has a variety of receptors to sense its environment in a wide spectrum.

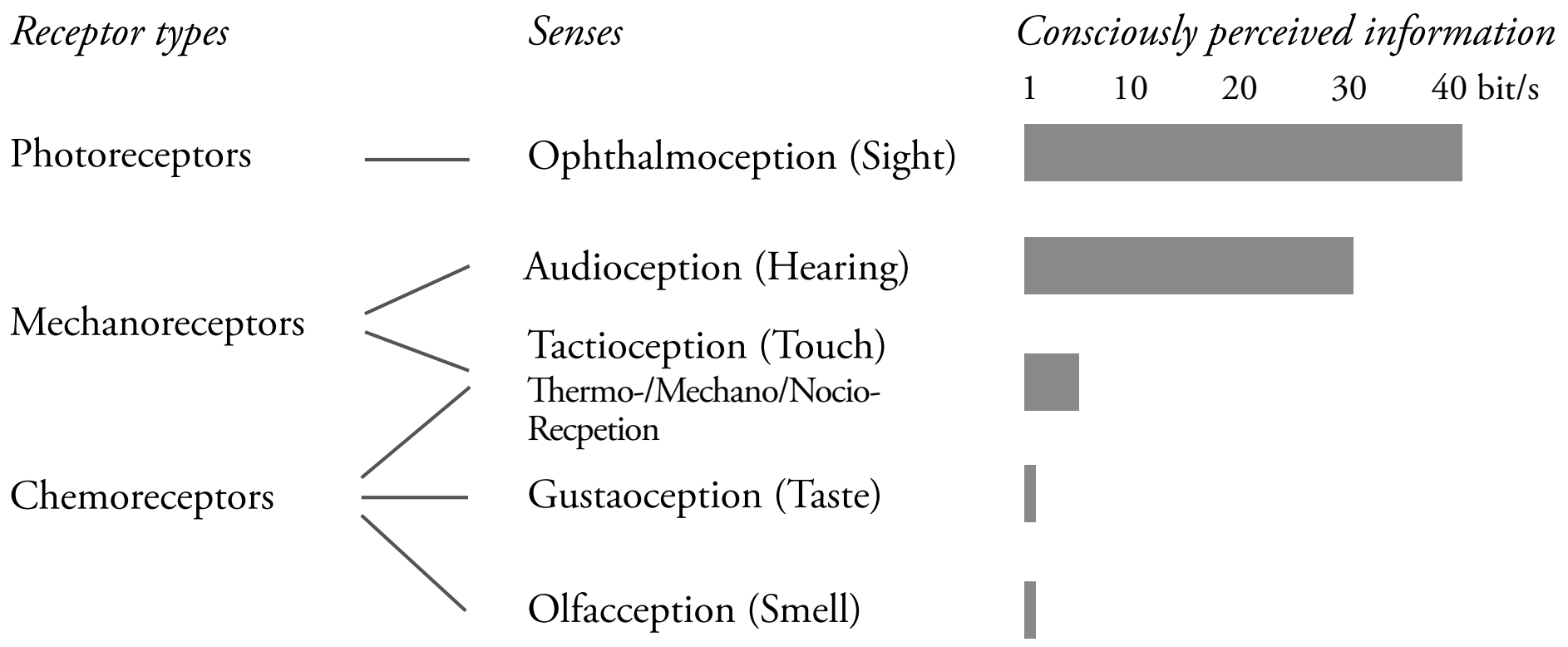

Figure 1. Showing the receptor types, their corresponding senses, and the fundamental capacity of each of these senses ("consciously perceived information" density based on Nørretranders (1998) [41]).

#### *2.1.1 Ophthalmoception (Sight).*

It becomes clear that the visual perception yields the highest bandwidth to perceive information. However, visual perception, like all senses, are subject to the availability of the wave they are detecting and they suffer from noise of adjacent waves of the same kind. Apparently, unfavorable light conditions can extremely interfere and limit the spectrum of our visual perception. Many ubiquitous computing systems, and thus notifications, primarily rely on the user's visual perception, which can cause a visual and consequently, a mental overload. While this is apparently negative, visual feedback can also affect the health of our bodies in a positive way as known from fundamental medical investigations. For instance, bright light potentially improves vitality and alleviates distress (Partonen and Lönnqvist, 2000 [42]). Moreover, it is known that adjusting these individually to the user's rhythm yields the power for aiding the body. For example, orange light waves can be described as visually bright, as it is considered to be warm, stimulating and moving. Dark orange light with a wavelength of 628 nm is generally perceived as being comfortable. Additionally, pulsating light can cause a quiet heartbeat, affects the brain wave activity, and thus the state of consciousness (Photosonix, 2015 [45]).

In HCI, visual notifications are broadly used with devices that incorporate a screen. Moreover, light has been used to create awareness while allowing the visualization of binary information such as an ongoing energy consumption (Tong et al., 2015 [56]) or ambient information [40]. Although a few mobile devices incorporate some ambient light notifications, it remains unclear as how best to incorporate an ambient light in wearables to convey different scales of notifications.

*2.1.2 Audioception (Hearing).* Although the density of the actual neuro-chemical receptors is comparably small, the perceivable information density is comparably high. What we know from medical research is that any kinds of sound, such as music or simple tunes, have a substantial impact (Gaver, 1993 [21]) on our physical condition. Following literature, musical stimuli can have an effect on our subjective perception of pain, on our heart rate, blood pressure, breathing rate, oxygen consumption, metabolism, and brain activity (Sawhney, N., and Schmandt, 1999 [53]). As we may have experienced on our own: unpleasant noise can also cause adverse mental state changes. Favored music instead, can be encouraging, inducing positivity and thus creating relaxation. It has been specifically proven that listening to music can create emotions such as joy and happiness right up to total intoxication (Schnell, 2005 [53]).

In HCI, auditory notification interfaces are common as they can be found everywhere (e.g., ringtone). In Virtual Reality (VR), audio effects also play an important role – such as improving immersion (DeGötzen et al., 2007 [13]). As an alternative to visual notifications, utilizing audioception looks promising in terms of information density. However, in public spaces audio is often not an appropriate channel for interactions, such as for the purpose of providing notifications, because our surroundings often contain high levels of noise. And headphones could isolate the user and thus might cause dangerous situations (e.g. a nearing motorcycle in traffic).

*2.1.3 Tactioception (Touch).* Tactioception pronounces Thermo- Mechano- and Nocio- Reception all over our bodies, thus we can feel heat, cold, pressure and any other sensation based on haptic touch. The term haptic technology was introduced in the late 1980s to define the aspects of human-machine touch interaction (Saddik et al., 2011 [17]). Currently, the term has brought together many different disciplines, including biomechanics, psychology, neurophysiology, engineering, and computer science, to refer to the study of human touch and force feedback with the external environment. Sensory cells on our skin interpret mechanical forces such as pressure, touch, vibration and strain into nerve impulses, called mechanoreceptors (Cholewiak et al, 2001 [11]; Zühlke, 2012 [65]).

In HCI, haptic interfaces used for notifications are generally divided into two different classes: tactile and kinaesthetic. The first one provides external stimuli essentially on the skin through a device. In contrast, kinaesthetic is related to stimulation of muscles, joints or tendons. Some typical kinaesthetic device configurations are manipulandums, grasps and exoskeletons. Another kinaesthetic interface is electrical muscle stimulation (EMS) such as that demonstrated by Lopes et al. (2015) [34]. In terms of haptic sensation, related work demonstrates vibrational (also denoted as vibtrotactile) feedback to be frequently used for providing notifications. This is because it is highly noticeable, since it yields a very unusual sensation Roumen et al. (2015) [49]. Having a look at use cases for haptic feedback, we can often find VR applications (Hayward et al., 2004 [23]) and navigation scenarios, in which alternative feedback is being applied (Meier et al., 2015 [39]).

Moreover, the ability to perceive temperature changes also belongs to the category of haptic sensation. In general, the perception of temperature is an individual phenomenon as the expression of heat and cold thermal receptors is not similar across users. In physiological treatment, heat stimuli are used to ease muscles (Prentice Jr, 1998 [47]). In contrast, cold stimuli can be beneficial to treat symptoms of exercise-induced muscle damage (Eston and Peters, 1999 [19]).

In HCI, notifications using thermal feedback can be applied in noisy and bumpy environments (Wilson et al., 2011 [63]), however, it is still not broadly being considered (the high energy consumption is a critical issue).

*2.1.4 Gustaoception (Taste).* Gustatory perception purely relies on chemoreceptors. The sensation is produced orally onto the tongue while chemical processes are triggered. In terms of information density, we consider it as rather low-resolute, although there are 2000-5000 taste buds located on the front and back of our tongue.

In HCI, we can also find taste interfaces, but which are not widely spread. While most interfaces apply chemicals, Ranasinghe et al. (2012) [48] introduced a non-invasive tongue interface based on electrical and thermal stimulation on the human tongue. Their results indicate that sour (strong), bitter (mild), and salty (mild) are the main sensations that can be simulated. While using gustatory perception for notifying may be interesting to explore, it is questionable whether, in the light of hygiene and safety issues taste interface will be pursued in the short term.

*2.1.5 Olfacception (Smell).* Olfactory perception is similar to the gustatory perception purely based on chemoreception. The process of smelling can be considered as a complex process, while a gaseous substance needs to be in contact with chemical molecules resting in the olfactory epithelium. Similar to taste, smelling yields a rather low bandwidth of information density.

In HCI, smell interfaces are similar to taste interfaces in that they have not been widely explored yet. Still, Henry et al. demonstrated a nose gesture interface that can simulate a particular smell to increase immersion in VR applications (Henry et al., 1991 [25]). However, it is conceivable to create scalable notifications based on the intensity and type of smell.

## 2.2 Forms and nuances of notifications

Presently, notifications appear in many forms, while being pronounced in different nuances. When it comes to mobile computing, focussing in particular on smartphones, most common notifications include visual pop-ups and accompanying vibrations. Vibrations especially appear to be highly noticeable (Roumen et al., 2015 [49]) in comparison to alternative feedback types, such as temperature, visual, and audio feedback. This may be the reason why vibration has grown to be a popular feedback type for mobile and wearable devices. A very common use case in research includes navigation aids for visually impaired users. However, eyes-free interaction can also be beneficial for healthy users such as when being on the move or when being occupied with other real-world tasks. Depending on the design, notifications yield the power to convey binary as well as minor complex information to the user in speedy way, with or without obstructing the user from his primary task.

As demonstrated in *Figure 2*, notifications can appear in different forms and nuances. While they can be binary, such as an LED being switched on, notifications can become complex, such as when increasing the dimension (e.g., 2D or multi-dimension array of LEDs), when using patterns (e.g., Morse-code blinking LED), and when varying intensity (e.g., brightness or colour of an LED). Combining all those attributes increases the complexity of a notification which, on the one hand enables increased density of the level of information, but, on the other hand, may be very attention drawing and thus likely to obstruct the user in his primary task, since cognitive load may be high. While this example demonstrates visual feedback, we can also transfer this to other feedback channels, such as audio feedback (dimension: channels; pattern: frequency; intensity: amplitude), tactile feedback by vibration motors (dimension: number of motors; pattern: sequences; intensity: power), etc.

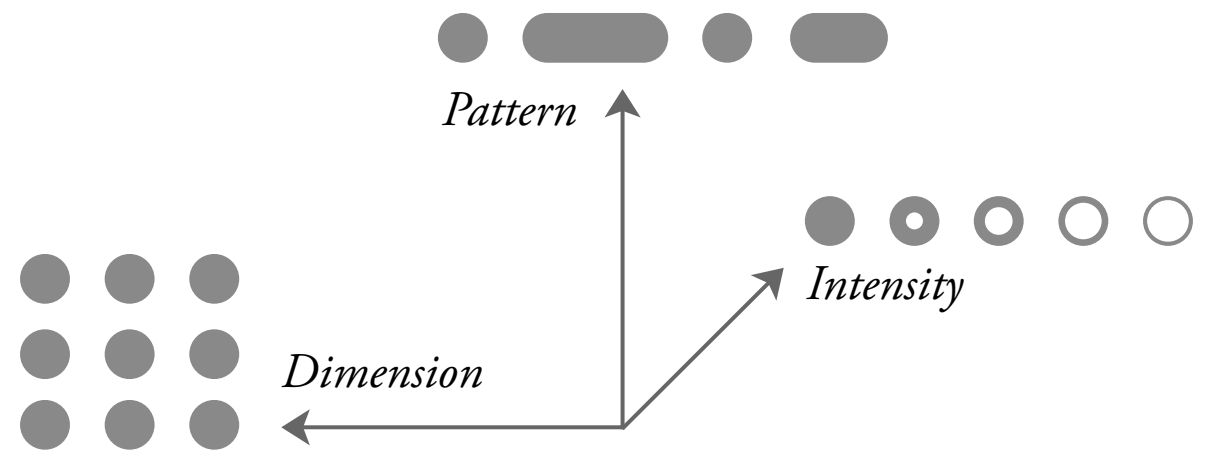

Figure 2. Illustrates the complexity of notification forms and nuances. Most common variations based on literature rely on dimension, patterns and intensity. Combining the different properties result in more complex notification designs.

Research presents a great variety of those variations that are based on vibrotactile notifications which, for instance can assist visually impaired people, such as for the use of pedestrian navigation. In the following, we present examples demonstrating all three directions: dimension, pattern, intensity.

*2.2.1 Dimension.* Feedback can be distributed spatially in a single- or multi-dimension. Feedback on a single dimension could be, for instance, a single tone, a single LED, or a single vibration motor. Furthermore, we can increase the quantity of actuators and thus the dimension. In research, we can find various examples for that, such as an array of vibration motors surrounding the user's wrist to provide notifications for spatial guidance (Weber et al., 2011 [61]), while only one vibration motor is actuated at once. A similar concept is proposed with *ActiveBelt* (Tsukada and Yasumura, 2004 [57]), a notification belt incorporating 8 vibration motors attached at a 45° angle, to accomplish navigation for pedestrians, when riding a bike (Steltenpohl and Bouwer, 2013 [55]), and when piloting a boat and flying a helicopter (Van Erp et al., 2005 [58]).

Thus, we can say, increasing the dimension, such as spatially distributing actuators, in particular vibrotactile actuators, can be beneficial when conveying binary and low-complex notifications, such as for spatial navigation.

*2.2.2 Pattern.* Conveying notifications beyond the complexity of binary stages, we can utilize temporal patterns. For instance, within a low dimension of a single actuator, such as using the vibration motor of a smartphone. *PocketNavigator* (Pielot et al., 2010 [46]) and *Navibration* (Badju et al., 2013 [5]) both propose exploiting such one-dimensional vibration patterns for a pedestrian navigation, while the smartphone rests in the user's pocket. Actions, such as directional cues are assigned to very specific vibration patterns, which need to be learned by the users. Other typical examples include the Morse-Alphabet (Carron, 1986 [9]), which is usually transmitted via visual or audio feedback. In contrast, a multi-dimension setup, such as with a two-dimensional array of vibration motors, allows the conveying of more complex notifications. However, arranging actuators in a 2-D space also allows the perception of spatial stroke patterns, which are easier to recall by the user. Such spatiotemporal vibrotactile patterns are investigated on different body parts, such as arm, palm, thigh and waist in *OmniVib* (Alvina et al., 2015 [2]). Similar works investigate multi-dimensional vibrotactile patterns on other body parts, such as on the foot (Meier et al., 2015 [39]; Velázquez et al., 2012 [59]).

Although notifications can be conveyed in many other ways, recent literature presents extensive investigations on vibrotactile patterns for notifications.

*2.2.3 Intensity.* Another possibility to expand the density of information is to adjust intensity of notification. By a change of intensity, we mean sounds to play louder, LEDs to shine brighter, vibration motors to actuate stronger, etc. For example, changing intensities of vibration motors has been demonstrated in *HaptiColor* (Carcedo et al., 2016 [8]), in which a wristband encodes colors into vibrotactile patterns. Other works that exhibit tactile feedback is presented by Huisman et al. (2016) [27] who attached a vibrotactile array on the inner side of the lower arm that generates different intensities of stroking touches in order to produce emotional responses. Intensity of sensation especially plays a role when applying thermal feedback, such as on the head (Peiris et al., 2016 [43]) and when trying to create emotional feedback, since users demonstrate different thresholds of sensibility (Arafsha et al., 2015 [3]). In parallel to haptics, we can of course also vary intensity for other types of feedback, such as taste interfaces (Ranasinghe et al., 2012 [48]).

While the human possesses a variety of senses, which are all set individually by nature, adjusting intensity individually, regardless of the type of feedback, may be an important adjustment screw to improve user experience with notifications.

## 3 SCALING NOTIFICATIONS BEYOND ALERTS

Notifications play a major role in mobile computing, while instantaneously providing us with information at any time. While this is certainly beneficial to the user, there are also drawbacks. One well-known problem is based on the user's attention resources, which are limited, but often demanded when notifications suddenly pop up in an inappropriate moment and carrying irrelevant content (Mehrotra et al., 2015 [38]). At this point an ongoing primary task, such as driving a car, may be interrupted, which results in potential danger. In other situations, we may perceive notifications to be annoying or highly disturbing, such as when lying down to sleep. In addition, certain user groups, such as elderly people, may have a different perception threshold and may overlook notifications quickly or being frightened by a sudden notification, such as an alarm.

*Considering context.* To produce suitable stimuli and to reduce interruptions, researchers propose considering context, such as environmental changes (Kern and Schiele, 2003 [28]), the user's emotional state (Liu, 2004 [32]), the user's activity level (Ho, J., and Intille, 2005 [26]), and all together in conjunction with the time of the day (Pejovic and Musolesi, 2014 [44]). For instance, in 2005 already, Ho and Intille (2005) [26] conducted a study in which they apply a wireless accelerometer to the user's leg to find out about their current activity. In conclusion, the authors propose the strategy to delay incoming notifications until the user initiates a task switch, which is often correlated with physical activity, such as a posture change from sitting to walking. Using context information, Mehrotra et al. (2015) [38] proposes to use a machine learning approach to infer on opportune moments for interrupting a task by notifications.

*Making it scalable.* Another approach to not constantly overwhelm the user with obtrusive notifications would be to scale notifications in accordance to their importance. A good example for that is demonstrated in *Tactful Calling* (Hemmert et al., 2009 [24]), in which a user is enabled to set the importance of a call by pressing the call-button with varying pressure. Depending on the callers input, the person called receives either a subtle or rather obtrusive call (silent/blinking, vibrating, or a loud audio notification). A similar approach for a scalable notification is demonstrated with audio text messages in *Nomadic Radio* (Sawhney and Shmandt, 1999 [52]), in which seven increasing levels of audio feedback (silence, ambient cues, auditory cues, message summary, preview, full body, and foreground rendering) are conveyed via a wearable speaker worn around the neck.

As already indicated in 1999 by Sawhney and Shmandt (1999) [52] notifications should be both *scalable* and *contextual*. Considering context can be important in order to coordinate a good point of time for notification, while we can scale the required obtrusiveness of a notification, depending on its priority and other situational factors.

This paper wants to build on these previous researches and elevate notifications to a further level, while expanding scalable notification to a new stage that forces the user to take action, in a manner of a *forcing feedback*. In this paper, we demonstrate our concept with scaling mechano-pressure, thermal, and electrical feedback. To our knowledge, how to scale these feedback modalities from a subtle notification up to a *forcing feedback* has not yet demonstrated. Please note: While *forcing feedback* can be described as an interaction concept, it should not be confused with *force feedback*, which is a particular feedback modality that provides a physical force, such as to a joystick.

### 3.1 Force feedback and forcing feedback

A few centuries ago, *force feedback* became very popular in modern aviation, because interfaces, such as the steering joystick, are becoming mechanically detached from the actual control unit. Because of the missing physical connection, information on the wind pressing against the rudder is being represented with simulated forces, which provides resistance to the pushing of a user. While this is a rather rudimentary example, many other types of *force feedback* have been explored, such as gaming joysticks and other handheld controllers. However, the most interesting *force feedback* was demonstrated in research some years ago: Pedro Lopes and his team researched *force feedback* in HCI applications using electrical muscle stimulation (EMS) while stimulating muscles with electrical impulses in order to make the muscles contract, which created a kinaesthetic *force feedback*. For instance, Lopes and Baudisch (2013) [33] presents a mobile phone running an airplane videogame, which makes the muscles contract involuntarily, so the user is made to tilt the device sideways. In trying to resist this triggered motion, the user may feel a pain force in his arm.

In contrast, in research, we can also find similar feedback, but which is still different and which we could denote as a *forcing feedback*. Instead of simply creating a force to resist, we can go a step further and force the user's body parts to move in a specific way or direction. A *forcing feedback* can also be demonstrated by electric muscle stimulation (EMS), such as shown by Lopes et al. (2015) [34] who forces body parts to move, such as hand movements driven by the computer. Hassan et al. (2017) [22] also applies EMS to the calf muscles in order to force a different foot posture, which significantly improves walking performance. While an EMS system can be considered to provide intrinsic feedback, we can also force the user's limbs to move by external forces. For instance, Chen et al. (2016) [10] presents a motion guidance sleeve, which generates subtle motion of the forearm based on "eight artificial muscles". Using stepper motors, fishing lines and elastic bands, the sleeve imitates the muscle contraction to drive the forearm to rotate instinctively. The illusion of external artificial muscles creating a pulling force and sensation is being conveyed. Another kind of a pulling force is demonstrated in *Pull-Navi* (Kojima et al., 2009 [29]), which propose earrings that are able to provide directions through haptic force. The developed prototype is apparently, cumbersome and incorporates a helmet, two servo-motors, and an external frame reaching to both ears. Pulling an ear forces the user to slightly turn his head to the corresponding direction and thus the user automatically walks in this direction. Another phenomenon, known as the hanger reflex, produces a fictitious force and involuntary rotation of the body using skin deformation. This haptic-induced force illusion is previously known to occur at the human head, which rotates unexpectedly when the frontal region and the opposing rear region are pressed using a wearable device. Kon et al. (2016) [30] investigated the hanger reflex on the waist using a U-shaped aluminium-ring, which moves around the body and subtly forces the user to take turns according to the position of the waist ring. Forcing the user to walk in a different direction using shoe interfaces was introduced by Frey (2007) [20]. The shoe prototype called *CabBoots* contains electro motors installed into the thick insoles of the shoes that are able to change the weight distribution within the sole of the shoe. This way the foot is slightly exposed to a subtle pulling force which would automatically drag the user in a predetermined direction. While rather rough movements can be forced, tiny reflexes can also be triggered as demonstrated by Dementyev and Holz (2017) [14]. In *DualBlink*, several types of feedback (light flashes, physical taps, and small puffs of air near the eye) are undertaken to force the user to blink eyes. Considering the effect, we can also denote that as a kind of *forcing feedback*.

In conclusion, *force feedback* mainly induces a haptic force which follows the aim of drawing the user's attention (e.g., a recoiling joystick to provide information on the windy environment or on the texture of the road) and making the user resist that force. In contrast, a *forcing feedback* doesn't necessarily rely on a haptic force as it has the aim of (un-)consciously making the user take involuntary action (e.g., making eyes blink, dragging feet to a certain direction for navigation).

## 3.2 Concept

To extend this related work, we propose scalable notifications that go beyond simply notifying the user, but forcing the user to take action. Providing notifications with different intensities can be useful in several situations, such as when being in a group of people and obtrusive notifications may bother others. While subtle notifications can help here, they may also be quickly overlooked or intentionally ignored without great effort and disruption. Therefore, we envisage these types of notification to be suitable for less important information. Moreover, there may be very urgent notifications, such as important information in order to keep ourselves safe and healthy. These notifications should even be a level beyond obtrusive, in a way that they cannot be overlooked and we are possibly forced to take action.

To illustrate the concept of *Scaling notifications beyond alerts*, we have chosen five stages, whereby the silent stage does not provide any feedback and whereby *forcing feedback* provides an extraordinary strong feedback, which may be unpleasant but still not harmful to the user. For instance, providing a high level of heat should be unpleasant, but not result in causing skin blisters. The number of levels between forcing and silent feedback have been evaluated in extensive pilot studies. In result, we established that with our hand-crafted user interfaces, most of the pilot study participants were capable of distinguishing between three different stages. More stages of feedback were substantially more difficult to distinguish. In order to not unreasonably increase complexity and to still maintain a reasonable difference in perception, we decided on five stages.

| ...feedback | *silent* | *subtle* | *moderate* | *obtrusive* | *forcing* |
|---|---|---|---|---|---|
| pressure force (Nm) | 0 | 0.3 | 0.9 | 1.5 | 1.8 |
| pressure frequency (Hz) | 0 | 2 | 2.5 | 3.3 | 5 |
| thermal temperature (°C) | 24 | 30 | 35 | 43 | 68 |
| electrical current (mA) | 0 | 10 | 12 | 20 | 22 |

Figure 3. We developed three prototypes providing mechano-pressure, thermal, and electrical feedback. We conducted three studies, in which we tested to distinguish between five stages of notifications. The table provides an overview of all important parameters, to enable replicability of our feedback. Please note: parameters for electrical feedback are user-dependent and need to be set individually. Further information can be found at the according section (Study 1-3).

With our work, called *Scaling notifications beyond alerts*, we want to extend previous research while combining scalable notifications with *forcing feedback* by mechano-pressure, thermal, and electrical feedback. To underline our concept, we developed three prototypes that demonstrate the feasibility in three different scenarios: (1) body posture correction, (2) hand-arm-vibration (HAV) overuse prevention, (3) car speed control, in which we apply several nuances of tactile notifications, ranging from subtle to obtrusive feedback, plus a silent feedback and a *forcing feedback* (*see Figure 3*). We evaluated three independent scenarios, in order to illustrate the broad applicability of our proposed concept. Although our selected scenarios are long way from being realistic, we see our lab studies as a means to validate the feasibility of this proof-of-concept in a safe laboratory environment.

## 4 SAFETY AND ETHICAL CONCERNS

Safety regulations are particularly strict in Germany, especially for clinical research studies, which have to follow the *Vocational Regulations of Physicians* (*„Berufsordnung der Ärztinnen und Ärzte M-V*[1]*"* in the state of Mecklenburg-Western Pomerania). While the passage "*III. Besondere medizinische Verfahren und Forschung - § 13-16*", defines conditions and protocols for medical research studies, these regulations do not apply to our type of studies following the ethics resolutions of the state of Mecklenburg-Western Pomerania. Nevertheless, to assure the participants safety, we requested a review from the local ethical review board[2]. The ethics review committee confirmed our study to have followed common study protocols, to be harmless for the participants, and thus they issued an official *Ethical Safety Certificate (see Supplementals).*

Although, our user study does not fulfil the characteristics of a medical study, we were nevertheless guided by the general guidelines for studies, which apply to any research study involving human participants and which are:

*(1) The principle of the protection of healthy subjects:* The subject's well-being has always been within the primary focus of our experiments. Actuator thresholds have been selected in a way that no temporary, nor permanent damage occurs on the subject's body. For instance, the *forcing feedback* would never cause backache, skin blisters, or muscle cramps.

*(2) The principle of voluntary consent:* If the subject was not feeling well during the study, the participant could indicate this, so we could immediately interrupt, or if necessary abort the study. The participants were not forced to complete the studies. For instance, when using EMS, the sensation is sometimes perceived as being unnaturally strange with some users and an uncomfortable feeling arose at some point. One subject aborted the study.

*(3) The principle of confidentiality:* All data was anonymized. At the end of the study, we explicitly asked some participants for permission to publish quotations. All subjects that have been asked allowed the publishing of their qualitative comments in an anonymized way.

*(4) The right to withdraw at any time:* During the study and up to the date of the publication, the subjects were allowed to revoke their voluntary participation and the test results obtained.

*(5) Right to compensation in the event of a claim:* The statutory regulations give all subjects the right to make claims for damages in the case of direct or proven late consequences. In accepting to volunteer in our study, the participants waive the right asserting any liability claim towards the study leader and study observer in the unlikely event of physical and mental damage occurring during and/or directly after the study. Any legal claims thus would have to be directed against the Fraunhofer Society. Obviously, our prototypes are handcrafted and did not undergo a certification process. Thus, it is in the spectrum of the possible that dysfunction may occur, although we tried our best to minimize the risk by extensive stress tests and by adding additional fuses.

It was our aim to generate the best user experience possible, although the subject may also have possibly been exposed to unpleasant stimuli, as we duly informed each participant. Further disturbances and risks are unknown. We did our utmost to ensure that participant health was not in danger and performed the study with the greatest of care and concern. None of the subjects were subsequently harmed.

---

[1] Berufsordnung der Ärztinnen und Ärzte M-V: http://www.aek-mv.de/upload/file/aerzte/Recht/Rechtsquellen/Berufsordnung_5_6_Aenderung.pdf (accessed: 13/01/2018)
[2] Ethikkomission Rostock: https://ethik.med.uni-rostock.de/ (accessed: 13/01/2018)

## 5 STUDY 1: BODY POSTURE CORRECTION (MECHANO-PRESSURE FEEDBACK)

In our first study, we investigate the concept of *Scaling notifications beyond alerts* for the example scenario of a posture correction at a computer workplace. We evaluated the experience of nuances of mechano-pressure notifications and whether they have an impact on sitting posture.

### 5.1 Motivation

Poor back posture while sitting or standing usually results in spine stress and thus discomfort or pain is experienced. This can lead to changes of tissue and bone, potentially resulting in spinal musculoskeletal disorders, such as bone spurs and intervertebral disc damage (Allen, 2000 [1]). Spinal problems usually result in back pain, which often become chronical and thus result in a loss of overall quality of life. Spinal issues are especially very costly for society in several ways, since treatment is expensive and lengthy, and the patient is usually too disabled to work a normal day. Poor posture is very common, as it is estimated that about 80% of all adults suffer back pain at least once in their life time, while 10% will experience a relapse (El-Sayed et al., 2011 [18]). Literature has shown that making users continuously aware of poor posture significantly reduces out-of-posture tendencies and encourages healthy spinal habits (Wong and Wong, 2008 [64]). Awareness can be created such as by triggering discomforting events, but which are usually tried to be avoided by the user (Shin et al., 2016 [54]). It is important to increase patient awareness of poor postures, so they can correct spinal curvature using their own back muscles instead of using external support (Wong and Wong, 2008 [64]). To increase the awareness of poor sitting posture, we want to provide notifications using mechano-pressure feedback.

### 5.2 Hypotheses

In order to evaluate the scalability of notifications, we designed a lab study in which we tested whether the users would be able to perceive the nuances in feedback of mechano-pressure feedback and whether different levels would have an impact on the user's sitting posture. We assume the following hypotheses:

**H1**: A sole mounting of bands, providing constant pressure, has a positive impact on the sitting posture.
**H2**: All notifications will have a positive impact on improving the users' posture into an upright position.
**H3**: The intervening notification will be different to other notifications and will force users to take action.

### 5.3 Apparatus

We developed a haptuator device (*see Figure 4*), working as a back piece that is mounted on the area of the thoracic vertebrae, which is the middle-upper part of the spine. To keep the device in place, we fixed the device in two ways, using a chest-band and using Velcro-tape, which is wrapped around the shoulders. A firm fixation is crucial in order to enable the best sensation to the user. A very loose mounting would not guarantee that the device would stay in place, while notifications may not be easily distinguishable.

In terms of technology, we used a powerful 12V servo-motor providing torque of approximately 160oz-in. The force is translated via a horizontal gear into a linear force that slides the metal bar from two 5V Solenoid haptuator coils (ZHO-0420L) towards the back. There two small metal bars are push themselves just next to the vertebrate, the corpus of the spine. Our pilot studies have shown this setup to result in a better sensation then pushing directly onto the middle of the vertebrate. We found it necessary to use the Solenoid haptuators, since they allow for a poking at higher frequencies than the motor is capable of.

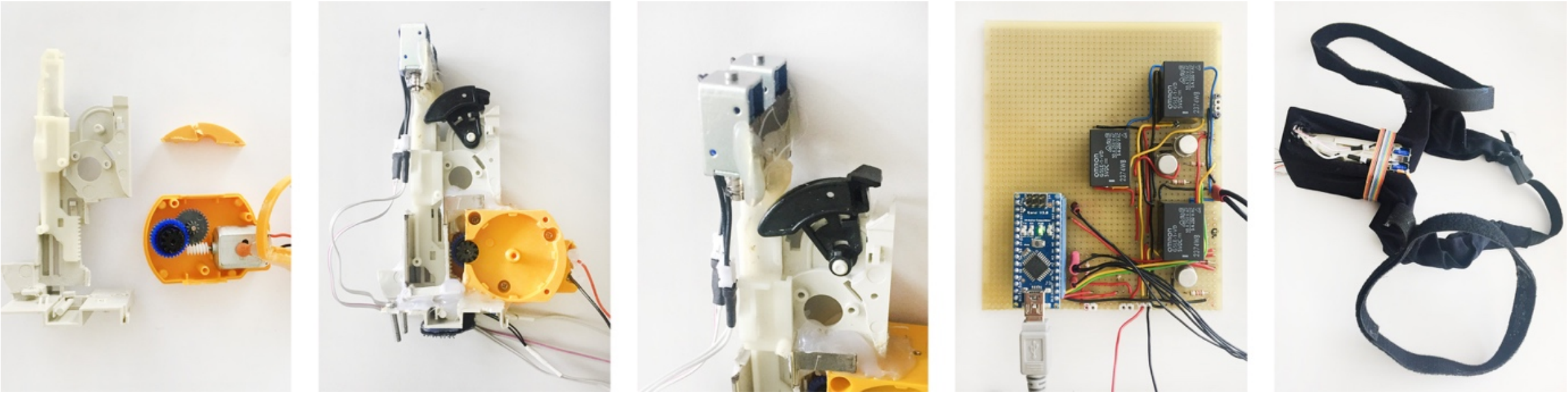

Figure 4. The left picture shows an early step of the development process. The rotational force of a 12V DC motor is transitioned into a sliding movement, while the motor is driven by a typical Motor Reversing Circuit using two Relays. Two 5V DC poking actuators are attached to the top enable for high frequency actuation, which are driven together by a single Relay. Since this is a custom hand-crafted prototype, we used hot glue to keep all parts in place. We implemented a capacitive sensor in order to know the position of the slider, while the DC motor as well as the poking actuators are driven by an Arduino Nano through three Relays (G5LE-1-VD). Since the prototype was capable of creating significant forces, we were required to mount the device in a stable manner by using a chest band and two arm/shoulder bands.

We set up our device to provide four nuances of notifications ranging from subtle feedback (Servo motor providing low pressure and the Solenoid haptuators poking in a low frequency) to *forcing feedback* (Servo motor providing maximum pressure and the Solenoid haputators poking in a high frequency) – *see also Figure 3*. Based on a pilot study, we decided to increase pressure and frequency at the same time to increase noticeability. Depending on the anatomic structure of the user's back, especially the nature of the costelas (ribs) and scapula (shoulder blades), the device was sometimes slightly less or more tightly attached to the spine, which can hinder the metal bars of the Solenoid haptuator to come out entirely.

## 5.4 Procedure

We instructed the user to sit in an upright position in front of a computer workstation. The face was approximately 100cm from the screen, while we made the position and angles of the user's arms to vary from approximately 45° to 130°. We provided the subject with a brief overview of the upcoming study, while the user was told to perceive tactile feedback at back. In order to prevent the user to produce an artificial behavior, we did not reveal the actual purpose of the study in beforehand but revealed it right after the study had been completed. At the introduction, the user orally provided information about his respective health status, in particular if they were aware of any chronic or acute spine disabilities. Moreover, the participant signed a consent letter confirming to voluntarily taking part in this study and not standing under influence of any drug. To assure safety, we followed general guidelines for studies (*see Chapter 4*). After the apparatus was mounted on the user's back, the study leader sat next to the workstation controlling the experiment, while remotely triggering the apparatus to actuate with five notifications (silent-, subtle-, moderate-, obtrusive-, forcing-feedback). Each of the five notifications was presented in a random order with a temporal displacement of 30s, then the keyboard was moved 10cm away from the subject (which resulted in an angle change of arms and back) and all five notifications were presented again in the same manner. In order to see whether the drift into a bad posture also occurs at different sitting angles, we tested different keyboard distances, while the last distance apparently forced the user to adopt a very poor sitting posture (bending over towards the screen). Since the user's posture worsened over time, we additionally repeated the whole experiment without providing any notifications.

## 5.5 Task

The task the users were asked to perform was simple, but highly demanding of the user's attention and thus very engaging. We occupied the user with playing a computer game called Slither.io, in which one is guiding a worm through a virtual world, by using the keyboard with both hands. Further task specifications have not been made, nor had the user been briefed broadly about the purpose of this study.

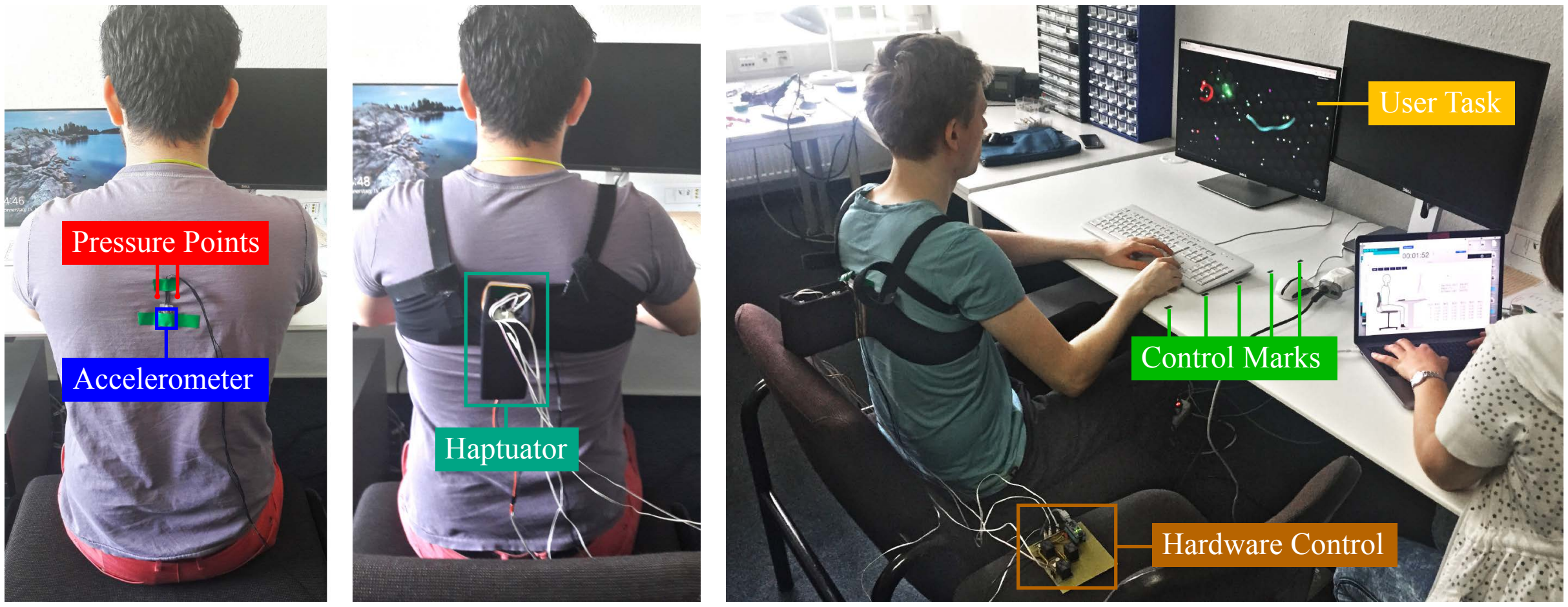

Figure 5. The left pictures show the attached accelerometer and the pressure points our haptuator is targeting after mounting. The right picture shows the study setup: the user playing Slither.io while the study leader is triggering the notifications. A self-build Java/Processing tool communicates with an Arduino Nano, while the user's posture is calculated in degree-angles by the accelerometer and dumped into a csv-file for post processing.

## 5.6 Methodology

Besides being very observant and noting down any qualitative feedback provided by the users, we mainly rely on quantitative data on the current sitting posture. For this, we attached an accelerometer to the user's back and translated the sensor raw data into degree-angles. We recorded the current angle of the spine just before and after providing the notification to the user's spine. We averaged the angle drift across all users to see whether notifications would help the users to correct their posture. The analysis was carried out with statistic methods to see whether the observed effects produced statistical significances.

## 5.7 Participants

We invited 13 users to take part in our study of which 10 participants were males. All subjects were students, mostly enrolled in computer science, aged from 20yrs to 35yrs (*M*=25yrs; *SD*=4.08yrs). Their body height ranged from 156cm to 183cm (*M*=175cm; *SD*=9.29cm) and their weight ranged from 48kg to 85kg (*M*=70.38kg; *SD*=12.74kg). The spectrum of our subjects reflects an average European citizen, although one of the subjects was outside this 'norm'; P4 – a female, 20yrs, 48kg, 156cm, however, is within the range of her body mass index (BMI) showing normal weight like every other subject except P3. According to the BMI, P3 – a male, 26yrs, 65kg, 183cm, indicates being slightly underweight and could therefore be seen as an outlier in terms of body mass. However, based on the study results, P3 performs similarly to other subjects. None of the subjects were aware of chronic or acute spine disabilities and considered themselves to be healthy.

## 5.8 Results

Our analysis is based on 975 data points, which represent the angles in degrees of the users' spine posture. At the beginning of our study, we recorded 325 data points = 13 users * 5 postures over time * 5 distances, without providing any notifications in order to see the posture drift occurring over time. (*see Figure 6 – left*). In the second part of our study, we recorded the users' posture change while having our prototype mounted to their back. Here, we recorded 650 data points = 13 users * 5 postures * 2 recordings (before & after triggering the notification) * 5 distances (*see Figure 6 – right*).

*Posture change over time (no notifications).* Sitting for longer periods of time has a negative influence on our back, since we inevitably change to a comfortable sitting posture that doesn't strain spine muscles, although this is not considered to be very healthy. This change of sitting posture happens unconsciously, while it can already be observed within the very first minutes after sitting down, as we can see in *Figure 6 – left.*

Just after we made the user sit down in an upright position (at a keyboard distance of 50cm, the angle of the back was something around 90°, depending on the user's spine and sensor position) we set a reference angle to 0°. After 10 seconds the users had already lowered their back by ø 1.23°. After 130s the users were already bending over for ø 4.64°. We then moved the keyboard 10cm farther away to a distance of 60cm. Here, we could determine the back posture to obviously worsen, while the back bent even more over to ø 6.21° and which further worsened after a time of 130s: ø 8.16°. We continued the experiment with different keyboard distances of 70cm, 80cm and 90cm to find out if this effect also occurs in other postures. As we can clearly see from the *Figure 6 – left*, the farther the keyboard is away, the crooked the back is. Also, the drift into an even worse posture continues.

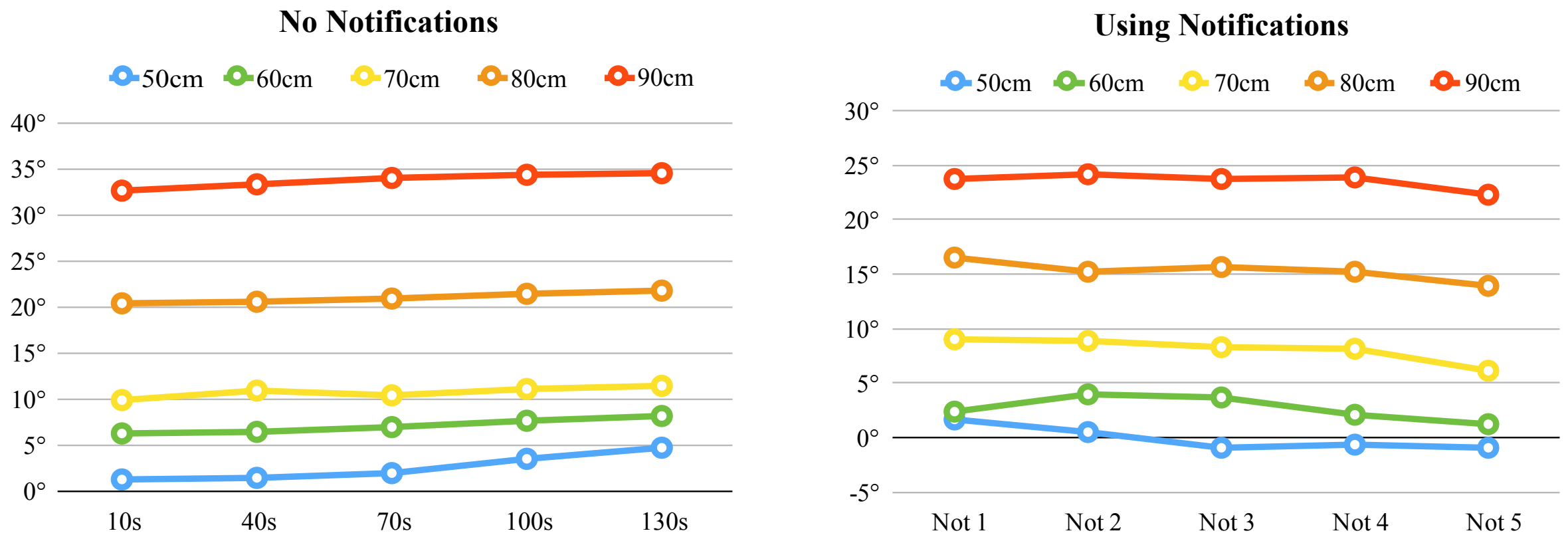

Figure 6. Both figures show the angle deviation in degree from the initial sitting posture. The left figure displays the time course of the posture change for 5 keyboard distances. The right figure shows current sitting angle after providing the user with notification 1-5. Note that notification 1 was silent and thus did not provide any tactile feedback.

This is also indicated by a *one-way ANOVA for correlated samples* ($F_{4,48}$= 68.73; $p$<.0001). A *Tukey HSD Test* could confirm almost all postures to be significantly different from each other and significantly poorer than the initial sitting posture. Furthermore, the postures shown at a keyboard distance of 60cm & 70cm do not seem to be significantly different, which is due to the relatively small sample size of $n$=13.

Next to this clear fact, we can now underline another finding with numbers: we can see a worsening of spine posture over time, while within 120s the back sinks into itself over all distances by ~2°. Comparing the average drift over time by an *ANOVA* doesn't show any significances ($F_{4,48}$=0.34; $p$=0.85), thus the strength of the drift describing the bending effect over time does not depend on the keyboard distance and will happen for any distance, at least for the very first 120s.

*Posture improvement (by mounting bands).* For our notification study, we mounted the haptuator prototype (*see Figure 5*) onto the user's back. We can confirm that solely having the mounting already creates a positive effect on the sitting posture, which is caused by the pressure created by the arm and chest mounting bands. The assumption, feeling the pressure of the mounting already improves user posture is confirmed by a *pairwise t-Test (correlated samples)* for each keyboard distance, such as for 50cm ($M_{no\ noti}$=2.5; $M_{noti}$=0.6; $t_5$=-2.91; $p$=.03), 60cm ($M_{no\ noti}$=7.11; $M_{noti}$=3; $t_5$=10.39; $p$<.0001), 70cm ($M_{no\ noti}$=10.65; $M_{noti}$=8.66; $t_5$=6.51; $p$=.0012), 80cm ($M_{no\ noti}$=21.05; $M_{noti}$=16.18; $t_5$=20.1; $p$<.0001), and 90cm ($M_{no\ noti}$=33.73; $M_{noti}$=24.31; $t_5$=39.54; $p$<.0001).

*Posture improvement (by notifications).* Although tightly mounted bands that push the shoulders back significantly improve sitting posture, we can still perceive a negative drift into a poor posture over time, but which is reduced from initially ~2° down to ~0.8° over all distances (within 130s per distance). However, when additionally provide tactile notifications onto the spine, the user's posture apparently improves (*see Figure 7*), which is also confirmed by statistical post hoc analysis.

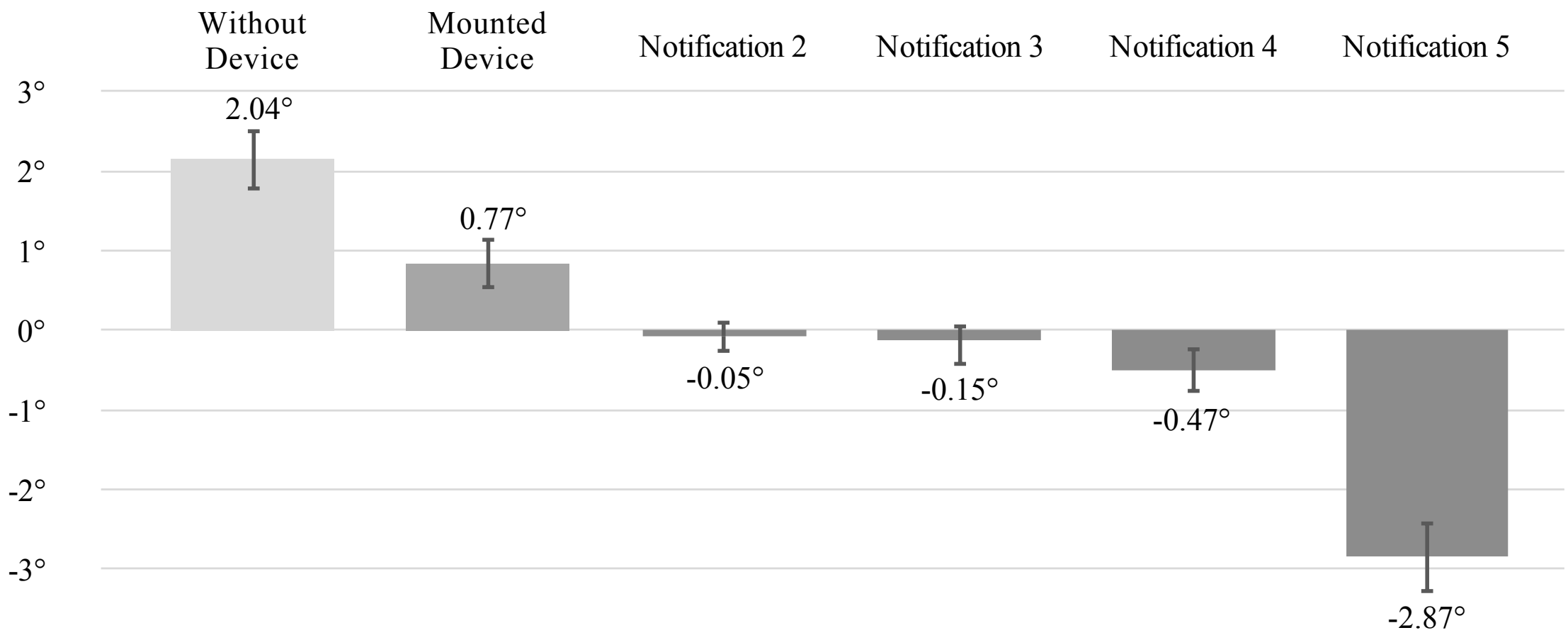

Figure 7. The diagram summarizes the average drift of the sitting posture the users show when having nothing mounted, having tight bands mounted to the back and shoulders and when additionally using tactile notifications (2-5). Note that positive angles represent a drift into a bad posture and negative values represent posture corrections.

A *one-way ANOVA for correlated samples* ($F_{2,8}$=29.72; $p$<.0002) shows that using notifications provide a statistical difference. A *Tukey HSD test* confirms the use of notifications to be strongly different from not applying anything ($M$=2.04°; $SD$=0.8°) as well as being different from mounting bands ($M$=0.77°; $SD$=1.04°). Therefore, we can conclude, that providing tactile notifications on the spine can significantly counter the natural drift into a poor sitting posture. While the average drift, sinking into a poor posture, is around ~2° over 120s, it looks suspicious that applying our haptuator would overcorrect posture, since we measured a correction of -2.87°. Therefore, we take a closer look at the notification stages. *Notification 1 – silent feedback / mounting* was clearly not providing any tactile feedback; therefore, we could see a drift of ~0.8° into a bad posture. *Notification 2* provided a slight tapping on the spine, but so subtle that it just marginally had an effect. While it is significantly different to "without device" ($t_4$=5.37; $p$=.006), it is not significantly different to just "mounted device" ($t_4$=1.75;

*p*=0.08). Although notifications 3-4 provide stronger feedback, which make the user correct their posture in a greater angle, the statistical outcome is similar to notification 2.

In contrast, notification 5 was significantly outstanding in every way. A *one-way ANOVA for correlated samples samples* ($F_{4,16}$=28.54; $p$<.0001) indicates notification 5 to be significantly different to any other notification, which is evidenced by a *Tukey HSD test*. Because our haptuator was heavily pushing just left and right next to the spine discs, the users were involuntarily forced to straighten their back.

## 5.9 Summary

Although this study needs to be seen as a proof of our proposed notification concept, we collected interesting insights into the scalability of mechano-pressure feedback in associated with sitting posture that now allows us to answer our assumed hypotheses (*see 5.2 Hypotheses*).

*Answering assumptions.* We can accept hypothesis 1 (**H1**), a tight arm/shoulder and chest band providing some pressure at the back already reduces that natural drift, although, it doesn't completely erase it. Still, this may already be good justification for wearing resistance bands such as those proposed by consumer products[3]. Even without instructing the users to straighten their back when receiving feedback (P8: *«I don't get the idea of why it is poking me»*), we established that providing subtle to obtrusive notifications at the spine even stopped the drift during and shortly after the feedback. This underlines our hypothesis 2 (**H2**), which can be accepted. However, an actual posture change into an upright posture is eventually needed to be performed consciously as also suggested by other consumer products[4,5].

Another way to significantly correct the user's back posture is forcing them by pressing with a greater force next to the spine discs, triggering a natural reflex that straightens the back. Following the data, after triggering notification 5, we are able to force the user to sit upright and to significantly change their back posture. Therefore, we suggest to also accept hypothesis 3 (**H3**), although we could perceive that two participants remained unmoved by *forcing feedback*. We assume this to occurred due to these users' individual perceptions. However, another reason could be due to a suboptimal mounting of our prototype.

*The individual factor.* Because of the nature of the costelas (ribs), scapula (shoulder blades), and the subjective perception on tactile sensitivity and pain threshold, users perceived the feedback in a weaker or stronger way. For example: some users enjoyed the *forcing feedback*: *«It felt a bit like a short back massage [...] the poking could be even harder»* (P2), while others immediately got in an upright position, pulling their shoulders back. P3 eventually started screaming: «S*top stop stop! Please, I am sitting straight again!*» In general, the study experiences varied greatly across users. One user (P6) was very much engaged with the game and stated: *«The noise produced by the motor and the relays are distracting me playing this game»*, while another user (P2) was bored by the game. However, the initial purpose of the game, which was in diverting the user's attention to something else other than on the haptuator, was achieved. Although the users' perception is very individual, we could show that our notification concept provides significant different nuances of notifications that were perceivable across all users.

---

[3] THE ERGO Posture Transformer: https://www.kickstarter.com/projects/708946960/the-ergo-posture-transformer-perfect-posture-insta (accessed: 13/01/2018)

[4] UPRIGHT GO: https://www.kickstarter.com/projects/upright-go/upright-go-fix-your-screen-slouch-correct-your-pos (accessed: 13/01/2018)

[5] Backbone: https://www.kickstarter.com/projects/gobackbone/backbone-the-smart-easy-way-to-a-healthy-back (accessed: 13/01/2018)

## 6 STUDY 2: HAV OVERDOSE PREVENTION (THERMAL FEEDBACK)

In this study, we investigate our proposed notification concept in the domain of work safety. We evaluated which nuances of notifications using thermal feedback would actually be recognizable on the palm of the hand when being exposed to hand-arm vibration.

### 6.1 Motivation

A lot of tools used by handcrafters or heavy workers emit considerable vibrations, which spread throughout the entire body. Due to the long-lasting and mostly intense vibration received by the hands and arms, irreparable damage may be caused to the sensorineural (Brammer et al., 1987 [6]) and muscular (Bovenzi et al., 1991 [7]) system. These diseases are well known and denoted as HAV- / Raynaud- / White Finger- Syndrome. In order to protect the workers, there are regulations, which demand an evaluation vibration exposure and to assess potential risks. For instance: The German Vibration Occupational Safety and Health Regulation[6], which is similar to the European regulations (Donati, 2008 [16]), obliges the employer to comply with the limit of the daily dosage of A(8) = 5 $m/s^2$ and to establish certain vibration reduction programs when exceeding a daily dose of A(8) = 2.5 $m/s^2$. Newer professional tools emitting considerable vibration already track the HAV exposure periods. Sensor-kits[7] can be attached to older tools, which are also commercially available. Alternatively, a manual evaluation has to be done, though this is obtrusive and interrupts workflows and obviously can be inaccurate. While in research, we can find several other prototypical solutions (Laput et al., 2015 [31]; Ward et al., 2006 [60]), HAV exposure dosage is also trackable by unmodified smartwatches (Matthies et al., 2016 [37]).

However, all these tracking systems have the same drawback: they do not inform the user in an adequate way about their current and daily HAV exposure dosage. For example, most devices provide numbers and graphs, which have to be looked up visually and which interrupts workflow. Regardless of their HAV exposure dose, reaching certain limits and even exceeding the daily limit is not communicated in a different way, although it would be important to the worker. We envisage a future notification system to unobtrusively provide the users with an idea of their current HAV exposure dose. Depending on the received level of vibration, we can provide adequate notifications which allow the user to take breaks at his will. Based on own interviews and a previous study (Matthies et al., 2016 [37]) it has shown that although the daily HAV dose is exceeded, many workers intentionally ignore this notification in order to get their work done. Therefore, it makes sense going beyond a simple notification and forcing the user to take a break through providing thermal feedback. In addition, when transitioning to thermal information instead of requiring the user's visual attention, we gain the clear advantage of reducing task interruptions.

### 6.2 Hypotheses

In order to test different nuances of notifications, we designed a lab study in which we tested whether the users would be able to perceive different levels of thermal feedback and whether different levels effect an impact on the task performance. We assume the following hypotheses:

**H4**: Subtle notifications are less recognizable and may be quickly overlooked.

**H5**: The intervening notification will force the user to interrupt the task.

**H6**: Except for an intervening notification, all other notifications will not negatively impact the task.

[6] Lärm- und Vibrations-Arbeitsschutzverordnung vom 6. März 2007. BGBI. I, S. 261: https://www.gesetze-im-internet.de/l_rmvibrationsarbschv/LärmVibrationsArbSchV.pdf (accessed: 13/01/2018)

[7] Castle Vexo H GA2006H Hand Arm Vibration Meter: https://www.castleshop.co.uk/ga2006h-vexo-hand-arm-vibration-meter.html (accessed: 13/01/2018)

## 6.3 Apparatus

According to literature, the most effective way to stimulate mechanoreceptors and chemoreceptors with thermal feedback is applying heat directly onto the skin. Since we are bound to have skin contact, there are limited options in terms of selecting a suitable spot of the body. We decided to apply different intensities of thermal notifications (*see Figure 3*) on the palm of the hand for several reasons:

*Receptor density.* It is fact that the density of receptive cells is highest on the hand compared to any other spot of the surface of the skin (Detjeen et al., 2005 [15]).

*Practicability.* Because we apply notifications based on heat, the sensation can quickly become unpleasant when heat exceeds the personal tolerance threshold. Therefore, we decided not to fix the actuator on the user, but instead, on the device the user holds in his hands. In a pilot study, we also experimented with a cold stimulus, however, creating a forcing feedback with cold was not feasible.

We selected a drilling machine and mounted a peltier element (TES1-127025) on its handle (*see Figure 8*). Additionally, we used modeling clay in order to smooth the sharp edges of the peltier element. When the device is carried, the peltier element was in direct contact with the palm. For control, we used an Arduino Nano with a relay control circuit to drive the peltier element, since it requires 12V and high electrical power up to 65W.

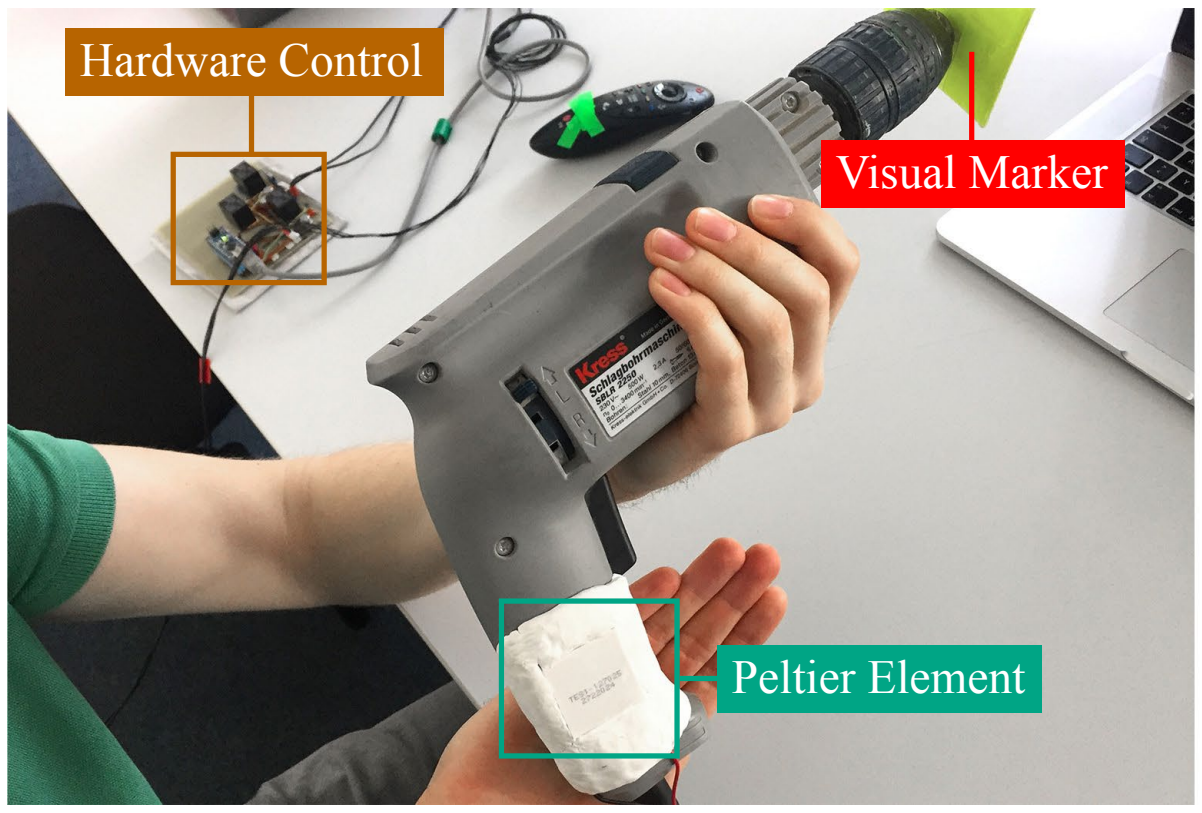

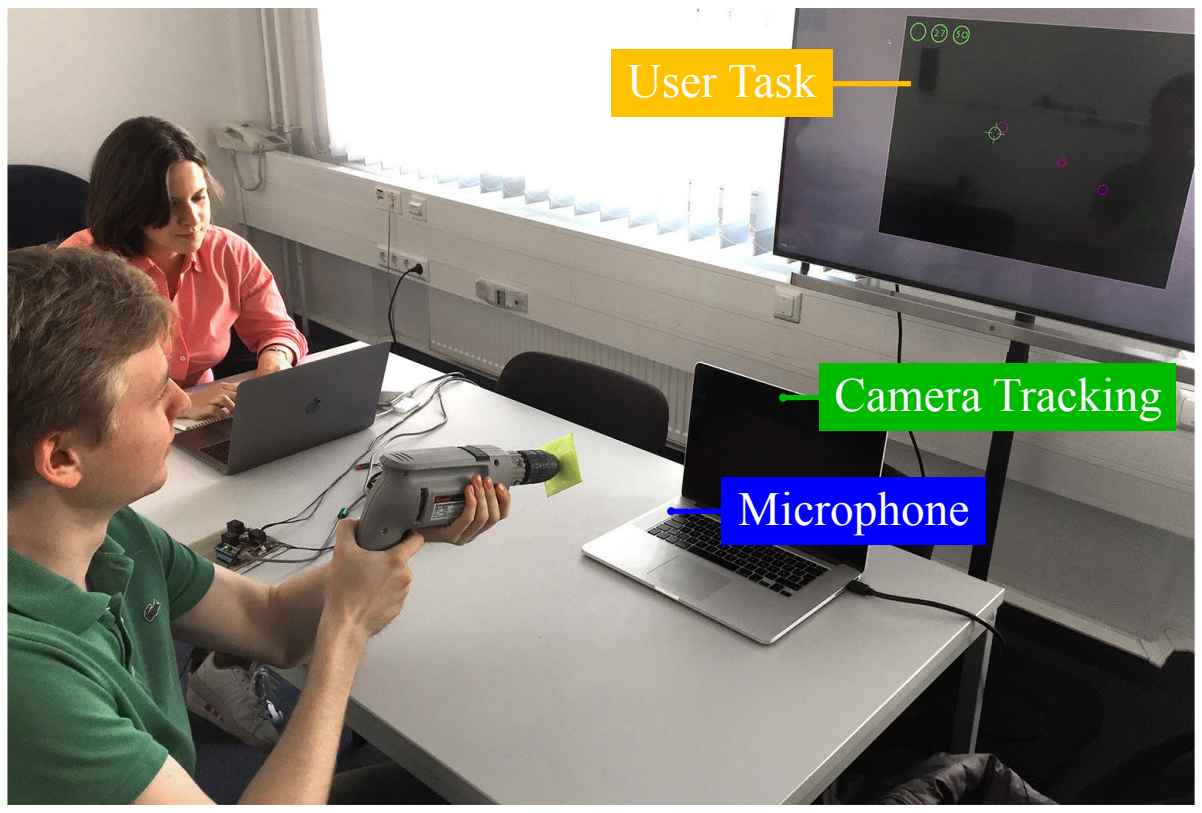

Figure 8. The left photo shows the prepared drilling machine, which has a peltier element (TES1-127025) and some modeling clay attached to the handle. The right photo shows the study setup; the subject had to aim at moving circles and drill them. The position tracking is done with an RGB web cam and the drilling is detected by the microphone.

In a prototyping manner, a piece of green colored paper was glued to the drill head (*see Figure 8*). A Macbook Pro web camera then tracked the current position of the drill, which was displayed as a crosshair in a Java/Processing application. While purple objects were moving in that application, the user was capable of aiming at them and drilling a hole in the air. The microphone was used in order to automatically recognize whether the user was drilling or pausing.

## 6.4 Procedure

We let the participant sit at a desk in front of a 50” LG Smart TV, which was approximately 1.5 meters away. The study leader was sitting diagonally across the subject and controlled the study (*see Figure 8*). After explaining the prototype and upcoming task in detail, we again asked the participant to sign a consent letter. Naturally, the safety of the subject had always been a priority *(see Chapter 4)*. All study participant had to explicitly state that they have been briefed, and that they voluntarily participate. After the introduction, the subject needed to pick up the drilling machine, while being required to hold it in the right hand. Usually the user made use of the left hand in order to support the quite heavy weight of the machine. There were three purple balls floating across the screen, which had to be aimed at and finally drilled. Drilling a hole takes 1.5 seconds, provided the user is within the hit box of the moving target. In order to prevent the user of non-stop-drilling, the drilling machine goes into a cool-down mode after a period of 3 seconds, which lasts another 3 seconds. Within that period, holes cannot be drilled.

While the subject were playing the game, in a random order each of the five notifications was presented for 10 seconds three times in a row, with a temporal displacement of 20s. The goal of this study design was to see how and when the user is capable of perceiving thermal feedback on the palm during a drilling task. Apparently, the increased cognitive load and the emitted vibration makes perception more difficult.

## 6.5 Task

We asked the study subject to work with the drilling machine, while the actual task was to drill "holes in the air", while the user was aiming the moving targets displayed on a 50" screen. The only goal was to drill as many holes as possible in order to increase the displayed score. Moreover, the user was asked to immediately tell the study leader as soon as he perceives thermal feedback on his palm. Nevertheless, we told the user they were allowed to drop the tool if they perceive an unbearable heat on the handle. No further task specifications have been made, nor had the user been briefed about the actual purpose of this study. Please note: even the highest heat would never exceed a level of 70°C and thus burnings, such as skin blisters, cannot occur. Moreover, the subject is only exposed to the maximum temperature for a few seconds and therefore, no irritations would occur.

## 6.6 Methodology

Also, in this study, we mainly rely on quantitative data. For each session, we recorded: (1) the reaction time from triggering until recognizing the notification, (2) the number of holes successfully drilled, and (3) the drilling time. For the reaction time, we rely on the user's oral response or on the physical response, such as putting the tool down. It is important to note that the reaction time is measured from the beginning of triggering the peltier element, until the user reports on feeling it. Therefore, our measured reaction time also includes a small amount of time, which is required to heat up the peltier element. Besides the qualitative data, we also noted down any qualitative feedback commented on by the users. The analysis was carried out with statistic methods to see whether the observed effects gave any statistical significances.

## 6.7 Participants

We invited 15 participants to take part in our study. The age of the participants ranged between 22yrs and 31yrs ($M$=25.5yrs; $SD$=3.11yrs). Their height ranged from 162cm up to a height of 189cm ($M$=178.73; $SD$=74.5cm), while their weight ranged between 46kg and 84kg ($M$=69.73; $SD$=11.3). The spectrum of our subjects reflects an average European, although one subject was outside this 'norm'; P13 – a female, 24yrs, 46kg, 162cm, can be considered as underweight according to the calculation of her body mass index (BMI). However, based on the study results, P13 performs similarly to other subjects. All other participants indicate being to be within their normal BMI. None of the participants suffered from HAV syndrome, while they considered themselves to be healthy.

## 6.8 Results

Notifications are perceivable in an individual way, such as the intensity and the pleasure or even pain. Because, our notifications range from (silent), subtle, moderate, obtrusive and forcing feedback, it may be that some of those notifications may not be recognized. We evaluated each notification level 3 times and thus base the following analysis on 180 data points: 15 participants * 3 trials * (5-1) notifications, since the first nuance of notification was subtle and thus did not affect the user in their performance.

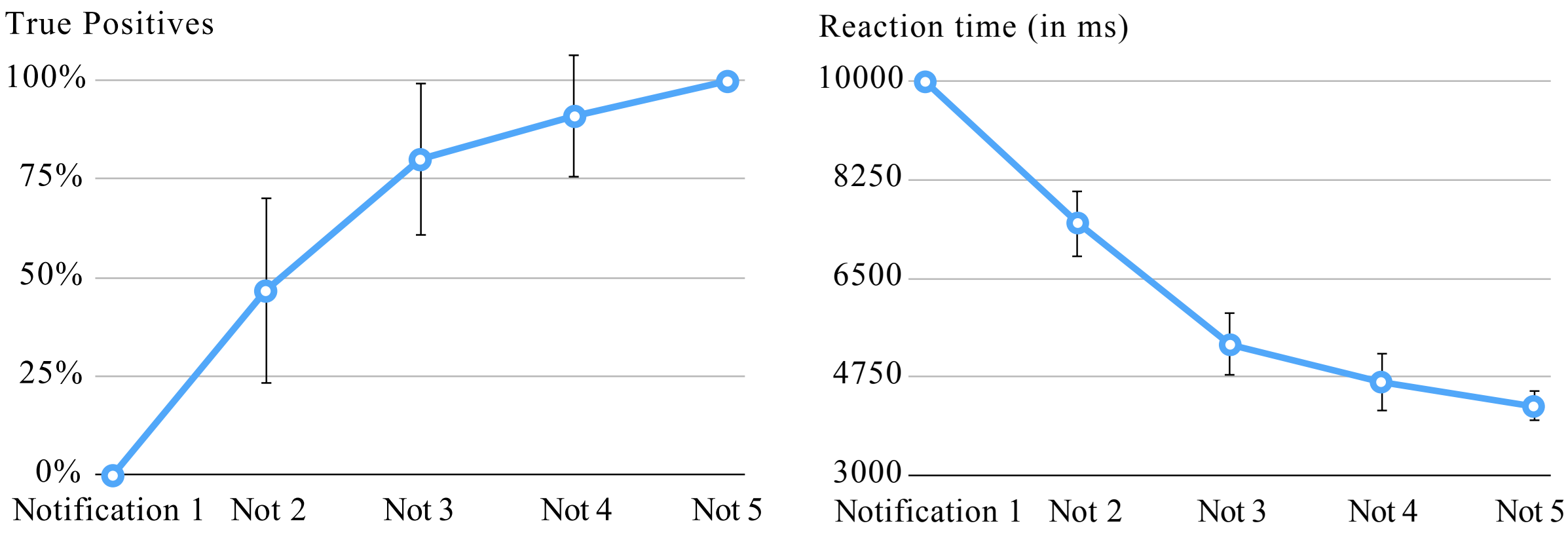

Figure 9. Left: The quantitative recognition of notifications among all users. Right: Reaction time per level of notification. (Error bars indicate the standard error)

*Recognition.* We ran a *one-way ANOVA for correlated samples* and found strong statistical differences ($F_{3,42}$=74.35; $p$<.0001) while checking which levels of notification have been recognized or missed (*Figure 9 - left*). A *Tukey HSD Test* reveals notification 1 to be significantly less recognized (*M*=0%) than any other notification, which is clear because our silent notification does not emit any heat. Notification 2 (subtle feedback) was perceived just around half of the time within the trials of all participants (*M*=46.6%; *SD*=30.34) and thus was perceived significantly different in comparison to all other notifications ($p$<.01). In terms of recognition, notification 3 (*M*=80%; *SD*=24.56%) and notification 4 (*M*=91.11; *SD*=19.79%) are not significantly different ($p$>.05) to each other. Both notifications, 3 and 4 have been significantly less recognized then notification 5. In contrast to all other notifications, notification 5, the *forcing feedback* was so alerting, that it could not be missed by any study subject (*M*=100%; *SD*=0%), which is significantly different than all other notifications ($p$<.01).

*Reaction time.* For each level of notification, we had three trials presenting a notification every 20s (10s actuation + 10s pause). When the user did not recognize the notification, we put a recognition time of 10s. We again, ran a *one-way ANOVA for correlated samples* and found strong significant differences ($F_{4,56}$=39.2; $p$<.0001); A *Tukey HSD Test* confirms notification 1 (*M*=1000ms; *SD*=0ms) to be statistically different ($p$<.01) to the other notifications, since notification 1 was silent and thus the user did not respond to it. In contrast, for notification 2 we provided a slight temperature increase up to 30°C, which was perceived on average after ~7s (*M*=7471.2ms; *SD*=2239.4). Notification 3 (*M*=4377.8ms; *SD*=1386.86ms), notification 4 (*M*=4407.27ms; *SD*=1782.87ms), and notification 5 (*M*=4228.13ms; *SD*=1014.17ms) are not statistically different to each other, but have all been recognized significantly faster ($p$<.01) than notifications 1 and 2 (*Figure 9 - right*).

*Average number of holes drilled.* While the user is perceiving notifications via the palm of his hand, it may be that his work performance is negatively affected. *A one-way ANOVA for correlated samples* ($F_{4,56}$=3.62; $p$=0.01) did evidence a significant performance drop. A *Tukey HSD Test* confirms notification 5 (*M*=2.95; *SD*=1.63) results in a decreased work performance, because the level of feedback was forced the users to put the tool down, since the heat was not bearable. No further differences were found; therefore, other nuances of notifications do not influence the performance negatively for our sample size (*Figure 10 – left*).

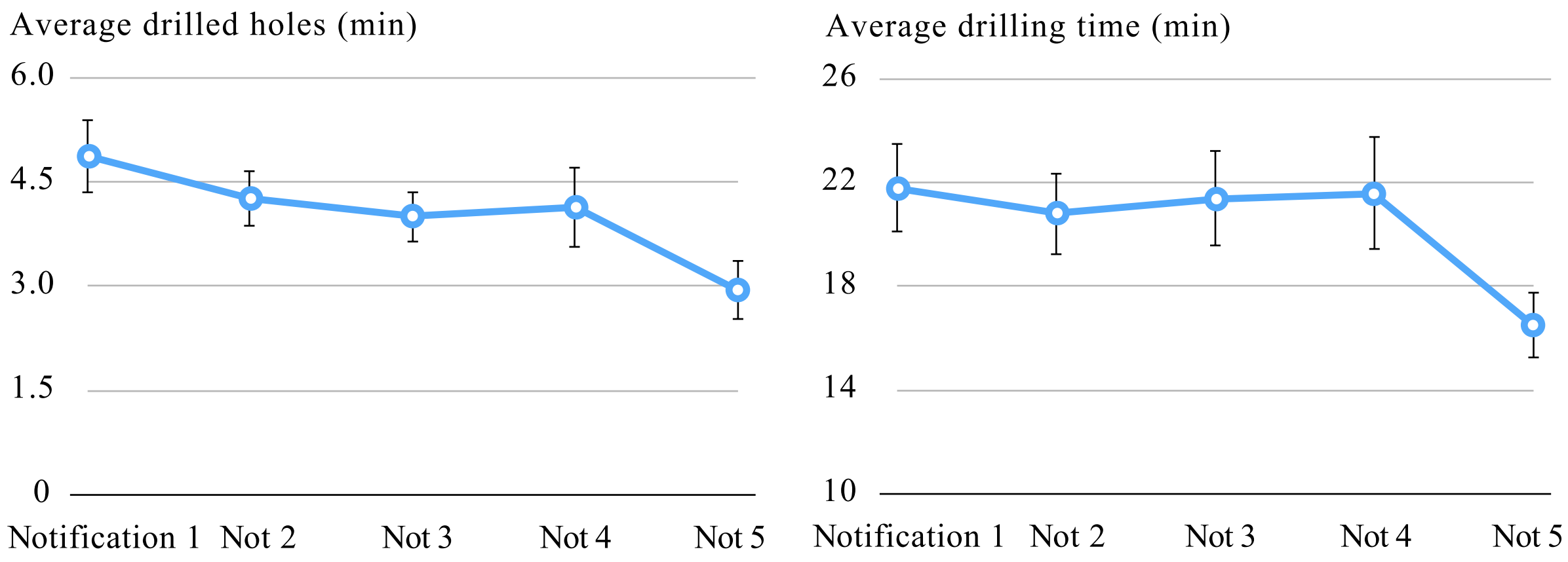

Figure 10. Left: The average number of holes drilled per minute among all users. Right: The average drilling time per minute among all users (Error bars indicate the standard error)

*Average drilling time.* We again ran a *one-way ANOVA for correlated samples* ($F_{4,56}$=4.56; *p*=.003) in order to check whether the notifications had an impact on the actual period of time the user drilled. A *Tukey HSD Test* revealed no significant differences between notifications 1 – 4. However, notification 5 (*M*=16.5s; *SD*=4.86s) demonstrates a significantly smaller amount of drilling time, which is clear because the drill was forcibly put down when notification 5 occurred and thus the user was disabled from using the tool (*Figure 10 – right*).

## 6.9 Summary

While we envisage *Scaling notifications beyond alerts* to handle the daily HAV doses of the tool, we obviously could not expose the user to considerable vibration for many hours in order to demonstrate such a system. Instead, we designed a lab study in a way that it would answer our assumed hypotheses (*see 6.2 Hypotheses*).

*Answering assumptions.* Hypothesis 4 (**H4**) can be *accepted.* The data analysis has shown that subtle notification levels are significantly less recognized than those which are obtrusive. Though, P13 stated notification 2 and 3 having a similar intensity. In general, qualitative feedback gathered from several subjects, confirmed thermal feedback to be perceivable in a very individual way. For instance, P8 stated *«Notification 4 is so hot – it is way too intense»*, however, similar to P10 the subjects did not put down the tool at notification 4. Most participants explicitly stated notification 5 to be too hot to touch, which was our goal – P12: *«Sorry, I can't take this, I have to put it down»*. Therefore, hypothesis 5 (**H5**) can be accepted. Each of our participants had to remove their right hand from the handle, which interrupted the task, when triggering notification 5. This is also indicated by a significantly lower averaged drilling time and by a significantly lower number of drilled holes. However, all other notifications (1-4) did not demonstrate any statistical differences in terms of task performance (number of drilled holes and drilling time). Hence, we can also *accept* hypothesis 6 (**H6**).

Regarding the drilling task, we generally had positive feedback, as the study itself was designed to be playful. However, we also received a negative comment from a young female – P8 stated: *«It is sometimes just too hard to hit the moving targets.* [...] *this is really frustrating»*. Another participant (P14), made another very interesting comment: *«Sorry, I was so concentrated on [recognizing] the heat, maybe I wasn't good at the shooting»*. However, his assumption contradicted the collected data, since he was the top scorer among all participants in terms of drilling performance, while he also recognized all notifications.

## 7 STUDY 3: CAR SPEED CONTROL (ELECTRICAL FEEDBACK)

In this last study, we apply our proposed notification concept to a more delicate domain; multi-tasking in a critical environment in which the user has to control a car while simultaneously interacting with the interior controls of the car. We evaluated different nuances of notifications and their effect on recognizing a changing speed limits.

### 7.1 Motivation

Driving a car is frequently mentioned when talking about dangerous daily tasks. Indeed, according to the U.S. Department of Transportation, motor vehicle crashes are still the leading cause of death for teenagers in the United States. In general, driving yields high risk, for experienced drivers also. Following statistical data, the worst driving hazards caused by the user are the consumption of alcohol (~11,000 fatalities /year), drowsiness (~5,500 fatalities /year) and cell phone use (~1,000 fatalities /year) in the United States[8]. It is clear that all these reasons are because of the inattention of the drivers. While driving should be the primary and only task, distraction occurs because of other tasks that are performed by drivers in parallel to the primary one. This can be critical when reaching a particular extent, such as when the primary task becomes the secondary task. This can happen quickly, when, for example, visually focusing on the interior controls or third-party devices, such as an iPod. This, in turn significantly affects driving performance such as keeping to the speed level (Salvucci et al., 2007 [51]) or keeping the car in the correct lane when being visually distracted (Chua et al., 2016 [12]).

We believe that notifications can help to redraw attention back to the driving task, especially when situations on the road change, such as the speed limit. Depending on the level of criticalness, notifications may reach from subtle towards obtrusive or may even force the user to take action immediately. To mimic a mobile scenario in a car, we designed a lab-based dual-task experiment with simulated driving as the primary task and an attention drawing secondary task, in which the user has to control different functions in the car's interior. However, we do not claim our setup to reflect real life driving, rather, it is meant to reflect a typical dual task scenario in which the secondary task competes for the user's attention and thus albeit unintentionally, becomes the primary task. To redraw attention to the driving task, we again provide different stages of notifications.

In this third study, we utilize an electrical stimulus, which we apply to the user's leg. Due to insurance reasons and safety regulations, we were not allowed to execute the study in a real car on the public road. Therefore, we designed a driving simulation in lab study, in which the user needs to control the speed of a car, while being distracted by secondary tasks, such as switching on a song on the radio.

The forcing notification would stimulate the muscle in a way that the foot is being involuntarily moved and hitting down the gas pedal. We agree that there are better ways to control a gas pedal, however, we just see this scenario as an example to demonstrate the power of a scaling notification using EMS.

### 7.2 Hypotheses

In order to demonstrate the feasibility of our concept *Scaling notifications beyond alerts*, we designed a lab study in which we tested whether the users would be able to perceive different levels of electrical feedback and whether different nuances of notifications have an impact on the task performance. We assume the following hypotheses:

**H7**: Subtle notifications are less recognizable and may be quickly overlooked.

**H8**: The intervening notification will force the car to accelerate significantly faster.

**H9**: Providing notifications with stimuli will improve the user's task performance.

[8] Autinsurance.org (Summary on driving hazards): http://www.autoinsurance.org/driving-hazards/ (accessed: 13/01/2018)

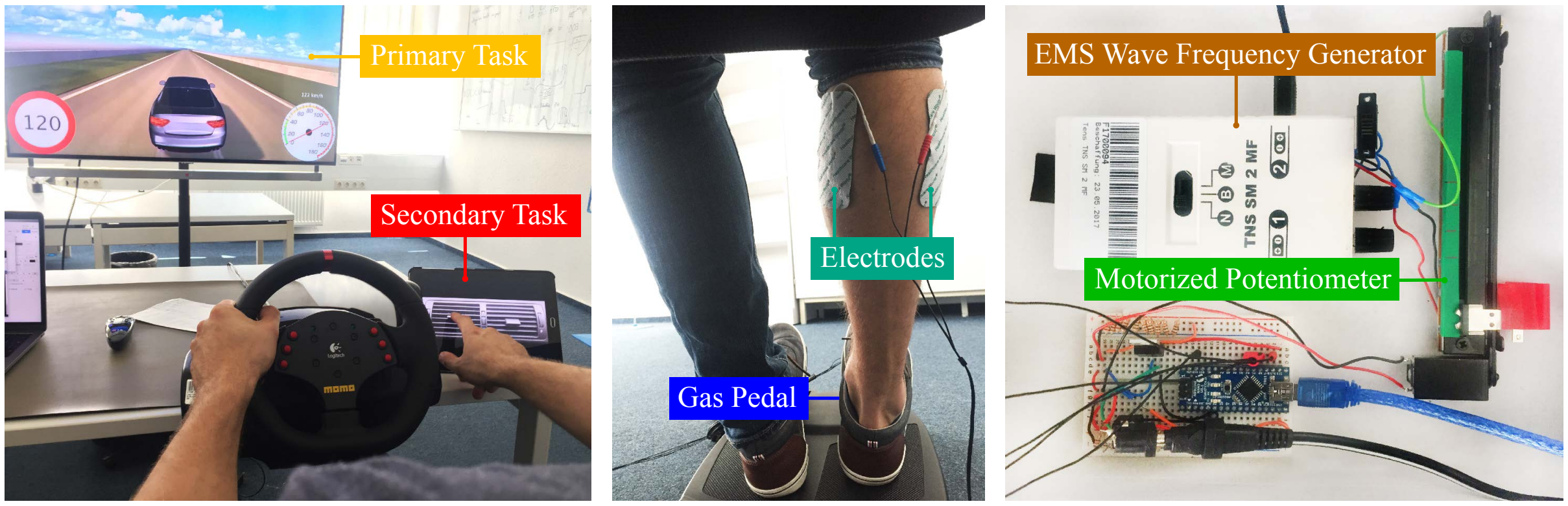

Figure 11. We designed a driving simulation, which is considered to be the primary task and simulated an interior on a tablet, which asked the user to perform a secondary task. While accelerating the car is done by pressing down the gas pedal with the right foot, we applied electrodes to the calf in order to provide notifications and control the foot movement. The EMS wave frequency generator was driven by our software using an Arduino Nano, which basically controlled the current knob that we replaced with a motorized potentiometer (10kOhm).

## 7.3 Apparatus

In order to provide automated notifications controlled by a computer application we used an Ardunio Nano and hacked an EMS wave frequency generator, also denoted as a nerve stimulator (Pierenkemper TNS SM 2 MF), which is capable of providing frequencies from 0.4Hz-100Hz. Our hack basically consisted of a hardware replacement of a knob by a 10kOhm motorized potentiometer (COM-10976) that is driven by an Arduino using a Power MOSFET circuit (2* IRF640). We used a single channel electrode mode, set a frequency of 80Hz, used a pulse width of 200µs, while we varied the output current from 0-70mA (*see Figure 11*). We designed 5 notifications which needed to be set individually per user. In a setup routine, we gradually increased the current floating through the leg-attached electrodes and asked when the user would feel a first light tingling. Then we continued increasing the current until the foot moved independently. The other stages between these two had been interpolated. *Figure 3* shows the averaged setting of current (in mA) across all users. The driving task was performed on a 50" LG Smart TV, while using a Thrustmaster Momo Racing steering wheel and foot pedals. The secondary task, interacting with car interior controls, was simulated on a Samsung Galaxy Tablet SM-T810.

## 7.4 Procedure

Before running the study, we introduced the apparatus and carefully described all aspects of the upcoming study. The participant then signed a consent form confirming to be informed about the study and possible risks. In particular when creating forcing feedback, not exceeding the threshold of an unpleasant stimuli, such as creating a painful experience at the user, is a risk we theoretically face. To erase this, we set up a user-dependent electric current threshold for the EMS stimulation, in order to create best user experience. We only varied the electric current. The type and amount of voltage as well as the position and size of the electrode remained the same across all subjects. Because the sensation is sometimes perceived unnaturally strange by some users, an uncomfortable feeling may arise at some point. Naturally, the participant had the chance to interrupt or to even abort the study, since they were all taking part voluntarily.

The participant had to sit at a desk in front of a 50" LG Smart TV, which was approximately 1.5 meters away. A driving simulation is displayed, which was controllable by a steering wheel mounted to the desk and gas pedals, which were placed under the table. A Samsung Tablet was placed on the right, just next to the driving wheel, which displayed different car interiors *(see Figure 12)*. The user was instructed to hold the steering wheel with at least one hand. The right hand was used to interact with the tablet, which switched between several interior screens. The user's feet were touching the gas pedal, which allowed them to change the speed. EMS skin surface gel electrodes were attached to the user's right calf, which received notifications by electrical stimulation of the musculus gastrocnemius caput mediale and laterale. A subtle notification creates an almost unnoticed tickling, the moderate notification creates a tingling of the complete calf, the obtrusive notification would tense the entire calf (feels like somebody clasps the leg tight), and the *forcing feedback* eventually causes the foot to press the pedal down. We chose this gesture, of pressing down the pedal to increase user experience of our study. Moving the foot upwards is technically-wise more complicated since we require more channels and more electrodes, which we would also require to be put underneath the foot, but which is however, impractical.

We tested all notifications while the user had the task of maintaining the car's speed, accelerating and slowing the car down. Each notification was tested 5 times in a row with a time disposal of 20s. The study leader, who was sitting next to the subject and controlling the study, had the chance to intervene during the study, in terms of delaying the notification when the users were not ready or aborting the study when the users no longer felt comfortable.

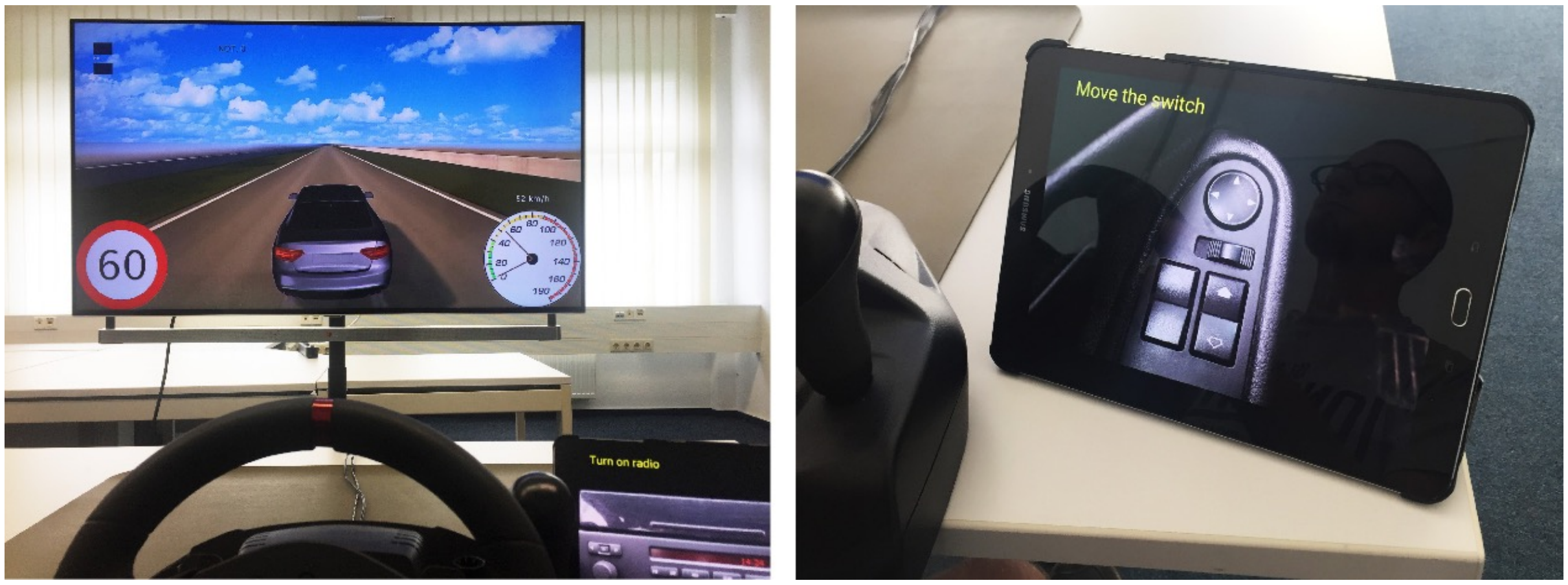

Figure 12. We built our own driving simulator, which showed the speed limit in the left corner, while a tachometer plus a digital number on the right side showed the current speed. On the tablet, we displayed different parts of a BMW interior, while a short request, colored in yellow, asked the user to perform an action.

## 7.5 Task

The user is occupied with two tasks: driving a car (*continuous* speed control by gas pedals), which is intended to be the primary task and interacting with interior controls displayed on a tablet (*discrete* input finger control), which is intended to be the secondary task.

The primary task, was to hold, increase or decrease speed according to the current speed limit traffic sign being displayed in the left corner of the screen. Changing the car's speed is done when pressing down the right gas pedal by the user's right foot. For the start of each round, we always asked the driver to keep a speed of 20kmh. Every 30-40s we started a new round in which the user had to change the speed. In one session, we provided 5 rounds with a speed change that required the driver to speed up the car by 20⟶40kmh, 20⟶60kmh, 20⟶80kmh, 20⟶100kmh, and 20⟶120kmh. The given speed changes were provided in a random order.

We conducted 6 sessions (including 5 rounds): while in the first session the driver is not distracted by a secondary task, and thus the user's attention can be fully devoted to the primary task. For the remaining five sessions, the user had to complete an additional secondary task on the tablet. The user was instructed to immediately complete the task appearing on the tablet, which was indicated by a "dring"- sound. For instance, the user had to switch on the radio, change the song, lower the temperature of a fan, switch on lights, etc. The task changed in a random order every 10 seconds.

While the users were exposed to a high level of distraction, we provided notifications by EMS to the user's leg as soon as a speed change occurs during the driving task. The user was instructed to immediately adapt to the suggested speed limit as soon as recognized. The additional notifications should help the user to recognize changes of the driving task. For each of the remaining 5 sessions, we tested a single nuance of notification.

## 7.6 Methodology

In order to investigate the differences of nuances of notifications, we have a look in which way they affect the primary task. Here we measured quantitative data of the driving task, which is: (1) reaction time, measured from the beginning of a changed speed limit sign until the user begins to significantly accelerate and (2) task completion time, which is the complete period of time needed to reach the new suggested speed limit. Moreover, the study leader needed to be very observant and note down significant qualitative feedback. This was especially important for this study, because the users may have rapidly felt uncomfortable due to the very unusual feedback modality. Because EMS works user-dependent, before starting the study, we evaluated the minimum current required to make the user feel a slight tingling and the maximum current that was required in order to make the foot move. Based on these values, we calculated the users' individual current for each notification. The averaged currents over all users are: $M_{Not2}$=10.7mA; $SD_{Not2}$=1.3mA; $M_{Not3}$=12.8mA; $SD_{Not3}$=1.5mA; $M_{Not4}$=20.1mA; $SD_{Not4}$=1.8mA; $M_{Not5}$=22.4mA; $SD_{Not5}$=2.1mA (*see also Figure 3*).

## 7.7 Participants

For this study, we asked the same participants to join for another session, who had already taken part in our previous study. Therefore, we had 15 participants aged between 22yrs and 31yrs (*M*=25.5yrs; *SD*=3.11yrs). Their height was ranged from 162cm up to a height of 189cm (*M*=178.73; *SD*=74.5cm), while their weight ranged between 46kg and 84kg (*M*=69.73; *SD*=11.3). All participants were European Citizens. In terms of body mass index (BMI), one of the subjects was outside this norm; P13 – a female, 24yrs, 46kg, 162cm, which can be considered as being underweight. All other participants indicated as being within their normal BMI. None of the participants had known muscle or nerve damage, such as Carpal Tunnel Syndrome, and considered themselves to be healthy. Only 11 out of 15 participants completed all tasks. P1, P2, P4, and P8 didn't complete the last notification (*forcing feedback*), with P1 already declining continuing the study during testing notification 4. The data of all completed trials have been included.

## 7.8 Results

In examining the reaction times (*Figure 13 – left*), the users showed when a change in speed limit occurred, there were significant differences when providing notifications, as evidenced by a *one-way ANOVA for correlated samples* ($F_{4,56}$=4.09; *p*=.0056). A *Tukey HSD Test* indicates notification 3 (*M*=1671.6ms; *SD*=555.91ms; *p*<.05), 4 (*M*=1600.4ms; *SD*= 552.99ms; *p*<.05) and 5 (*M*=1499.47ms; *SD*=305.34ms; *p*<.01) to provide better reaction times than using no feedback – notification 1 (*M*=2302.2ms; *SD*=977.78ms). Due to the rather small sample size, we cannot find a significant difference between notification 2 (*M*=1741.67ms; *SD*=591.62ms) and notification 1. It is assumed that the tasks occupied the users so much, that subtle feedback was sometimes overlooked.

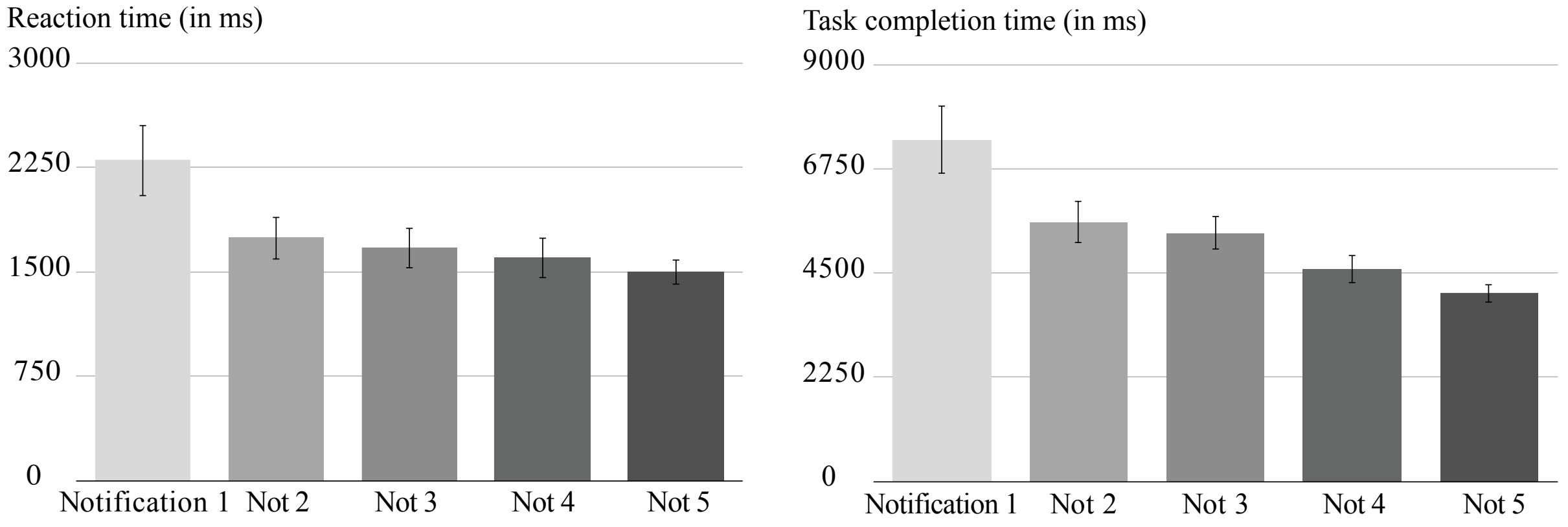

Figure 13. Left: Average reaction time for each notification. Right: Task completion time for each notification. (Error bars indicate the standard error)

Comparing the task completion time (*Figure 13 – right*), namely the time the user required to react plus reaching the given speed limit also gives significant differences following a *one-way ANOVA for correlated samples* ($F_{4,56}$=10.89; $p$<.0001). A *Tukey HSD Test* evidences notification 2 ($M$=5610.93ms; $SD$=1731.34ms; $p$<.05), 3 ($M$=5374.47ms; $SD$=1327.67ms; $p$<.01), 4 ($M$=4586.53ms; $SD$=1150.09ms; $p$<.01), and 5 ($M$=4073.27ms; $SD$=676.3ms; $p$<.01), helps the user to complete the task significantly faster than having no feedback – notification 1 ($M$=7385.47; $SD$=2826.16). No other significances found.

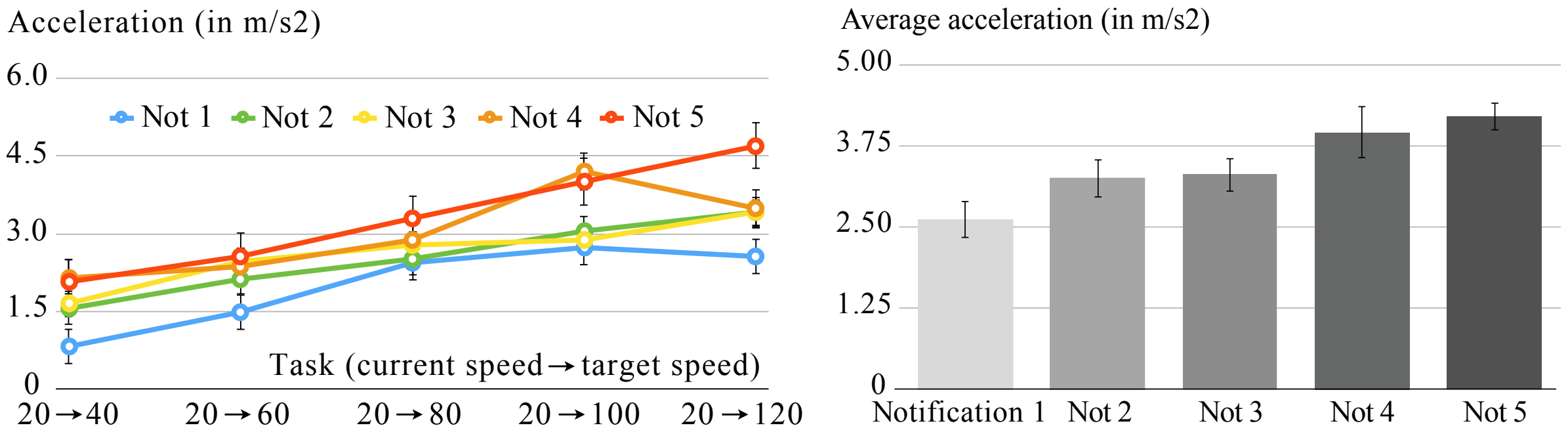

Figure 14. Left: Average acceleration of each task for each notification level. Right: Averaged acceleration for each notification. (Error bars indicate the standard error)

We also checked the average acceleration the users showed for each notification. From Figure 14 - left, it is clear that a higher speed limit made the user intuitively accelerate the car in a faster manner.

However, what is more interesting is the average acceleration occurring for each notification. From Figure 14 – right, we can infer that the user is deemed to accelerate increasingly quicker when receiving increased feedback. While examining the data, a *one-way ANOVA for correlated samples* ($F_{4,56}$=6.53; $p$<.0001) found strong significant differences, which is confirmed by a *Tukey HSD Test*. Increasing the obtrusiveness of notification, such as evidenced by notification 4 ($M$= 3.97m/s$^2$; $SD$=1.53m/s$^2$; $p$>.01) and notification 5 ($M$= 4.21m/s$^2$; $SD$=1.5m/s$^2$, $p$>.01) coerces the user to accelerate quicker in comparison to not having any feedback – notification 1 ($M$= 2.62m/s$^2$; $SD$=1.1m/s$^2$). Further differences could not be found with this comparably marginal sample size.

### 7.9 Summary

In this last study, we attempted to convey our notification concept to a driving scenario, in which the user is occupied with multiple tasks (controlling the speed of a car and interacting with the interior of the car). We do not claim to mirror a realistic driving experience, instead, we designed the lab study in order to see how our notification concept works with EMS feedback (*see 7.2 Hypotheses*).

*Answering assumptions.* Hypothesis 7 (**H7**) needs to be *rejected*. Although we overloaded the user with multiple tasks, the subtle feedback on his calf was always been recognized. We suspect this to be due to the very uncommon type of stimulus. Therefore, neither the reaction time, the task response time, nor the acceleration is significantly different when supplying subtle feedback or more obtrusive feedback. Providing notification 5, forces the user's foot to press the gas pedal down immediately until the end. The user was unable to gently accelerate and thus the acceleration at the points of the speed changes were significantly for notification 5 following the data analysis. Therefore, hypothesis 8 (**H8**) can be *accepted*, which is further confirmed by the user statements. P15 stated: *«It is hard to control the speed level with notification 5»*, P12 said *«I must fight to control my foot, because it is just moving»* and P7 explained: *«I wanted to stop at 60, but I couldn't control my foot»*. P7 also suggested this behavior made particular sense when hitting the break instead of the gas pedal. Regarding a task improvement when providing notifications, we can state that providing EMS feedback on the user's leg improves the awareness of the primary task. Providing notifications enables all users to complete the task significantly faster, while also the reaction time is substantially shorter. Therefore, we can *accept* hypothesis 9 (**H9**).

*The individual factor.* This study provided us with some interesting findings regarding the subjective perception of EMS notifications. Some participants perceived EMS as strong pain (P1), however, some of the other participants (P3) indeed enjoyed it. This particular subject also suggested that notification 5 should act as an emergency stop. While one participant (P10) explicitly stated being able to consciously distinguish between all nuances of notifications, however, we believe that distinguishing a scaled notification based on electrical feedback may not always be clear, also due to the individual threshold of perception. Therefore, we ran a calibration process before running the study. The setup of applying gel electrodes to the calf just before starting the study made some subjects suspicious. Three participants (P8,10,14) speculated that the electrodes may provide painful electro-shocks when exceeding the giving speed limit.

## 8 DISCUSSION

The idea of scaling notifications up to an extent that it is not just alerting, but forcing the user to take action is the key aspect of our concept. To make the perception differences between no notifications, scaled notifications and forcing ones visible, participants had to undergo three studies, in which they were stimulated with different feedback modalities with varying intensities. While our findings confirm perception to be individual, such as subtle feedback being sometimes overlooked, other test subjects demonstrated being very sensitive. Therefore, in future, it could be a challenge to find a subtle notification appropriate for a large number of users, based on their individual threshold of perception and based on contextual factors, such as environmental noise. However, we envisage a future notification system to also be capable of scaling up feedback based on contextual information (e.g., user's activity level, emotional state, and environmental changes). For instance, being in a meeting room, feedback may be subtle or silent in comparison to when being in a cafeteria. If working in a factory, notifications may strongly adapt to the urgency of a working task or safety monitoring. One could imagine that a system would notify the user when potential danger occurs, in an obtrusive manner, while transitioning over to a *forcing feedback* to protect the worker.

Also, because the threshold of perception is individual, we can clearly see that the transition from commonly used levels of feedback to *forcing feedback* is seamless. That is why making a clear cut and denoting *forcing feedback* to not being a type of notification may be questionable from our point of view. We see *forcing feedback* to be an extension of notification, because the user indeed is being made aware of a system's state change, plus an actuation created at the user's side. This mechanism of a *forcing notification* clearly brings up ethical concerns and discussions on the role of notifications. How far are notifications allowed to direct the user? Should we better create persuasion and incentives before forcing the user to take action? Is there any better way to guide the user to execute the desired action, instead of forcing them? We believe that scaling notifications up to an obtrusive level may be the necessary step to create awareness before taking over control. Moreover, creating persuasion and incentives can be done on a contextual level, such as by making use of gamification elements with scaling notifications before *forcing feedback* becomes necessary. Nevertheless, this is not the focus of our work; we aimed to demonstrate the technical feasibility of scalable notifications, which incorporate a *forcing feedback* that we see as the extreme pole of notifications.

In order to underline our theoretical concept, we conducted three lab studies, which on the one hand give an idea of the potential usefulness, although on the other hand they are not representative of real-world scenarios. Due to practicability and mainly safety reasons, we could not go to a real shop floor, nor make the subjects drive a real car. In a controlled environment, occurring effects are better recognizable and measurable, while potentially critical situations, such as when being involved in traffic and the user gets confused by *forcing feedback*, do not arise. Indeed, conflicts also occurred with the user's desire, such as to stop speeding the car up, although the foot was still pressing down on the gas pedal. This is especially crucial when it comes to a real-world scenario. It is questionable when do we really need a forcing level? Who is taking responsibility? What type of scenarios are actually suitable? These and many other questions now arise, but which cannot be answered sufficiently in a paper solely focused on demonstrating the technical feasibility based on lab studies. Furthermore, it is evident that the results gathered from a lab study may not be completely transferable to real world applications. In the real world, we have to deal with environmental noise and with the user's individual level of perception, which slightly varies over the day. Moreover, social acceptance of such interfaces may not be appropriate yet and thus the user may sometimes feel uncomfortable using such system. For instance, only two thirds of our test subjects (8/12 users) stated that they would be willing to adopt these methods of notification in a future system (choices: yes/no), although, the test subjects rated the perception of our studies to be neither uncomfortable, nor pleasant (3.3/5 points, based on a 5-point Likert-scale). The most preferred scenario was the drilling using the thermal feedback (6/12 users) in contrast to the haptuator (3/12 users) and to the EMS device (3/12 users).

Nevertheless, we believe the general concept to be tremendously relevant in future while being applicable in distinct scenarios, such as for occupational safety applications (e.g., to support the worker and preventing executing potentially dangerous tasks, while helping someone to avoid touching a particular area) or for games, which usually take place in a static environment, so risks are substantially reduced (e.g., a player interacting in a VR and using multiple modalities to support multitasking). We clearly see this paper as an early impetus for future work, which may focus on applying seamless scalable levels of notifications in a real-world scenario while considering context variables, such as the user's state or environmental influences.

## 9 CONCLUSION

In this paper, we introduced the concept of *Scaling notifications beyond alerts*, which expands on previous work on notifications, while we demonstrate scaling a single feedback type until exceeding the threshold of obtrusiveness and thus forcing the user to take action. While our lab studies may be a long way from being realistic and practical, they could nevertheless, emphasize our concept of working in a lab environment. Be it mechano-pressure, thermal, or electrical feedback, in each of our studies we were able to present scalable nuances of notifications, which demonstrated different effects on the task. We could also scale each type of feedback to an extent that it could not be ignored by the users, while forcing them to execute our desired action, such as sitting upright, pausing the drilling and pressing down the gas pedal.

We envisage our proposed notification concept, including the system to force an intervention, to provide a significant benefit when implemented into real applications, although further studies would be required. Overall, we believe scalable notification to be tremendously powerful, when considering context, such as the user's mental state, the environment and temporal variables. Scaling notifications in accordance to these conditions can enable the user to be less distracted and interrupted by scaling notifications to a subtle but still recognizable level. Moreover, the amount of overlooked notifications can be reduced, such as when scaling them to an obtrusive level in situations when high level of noise is present. In particular, an intervening notification can provide unique situational benefits, such as avoiding critical events (e.g., touching a hotplate, dangerous limp and spine postures, overregulating a control knob, crossing red lights, etc…). No matter what the application may look like, in future we need to have an ethical debate between the user's freedom to act and the system forced intervention and thus also about possible consequences and responsibilities.

## Authors' Vitae:

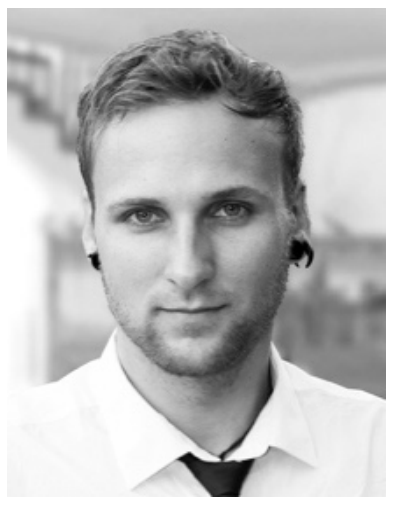

**Dr.-Ing. Denys J.C. Matthies, MSc., BA.**

Denys Matthies is a postdoctoral research fellow at the Augmented Human Lab, Auckland Bioengineering Institute, New Zealand. Before, he was a research associate at the Fraunhofer IGD Rostock, Germany. In 2012, he received his BA. degree in Interface Design from the University of Applied Sciences in Potsdam. In 2014, he received his MSc. degree in Human-Computer Interaction from the University of Munich. In 2018, he defended his PhD. (Dr.-Ing.) at the University of Rostock. In his professional career, Denys mainly worked in academia, such as at the Deutsche Telekom Innovation Laboratories in Berlin and at the HCI-Lab at the National University of Singapore.

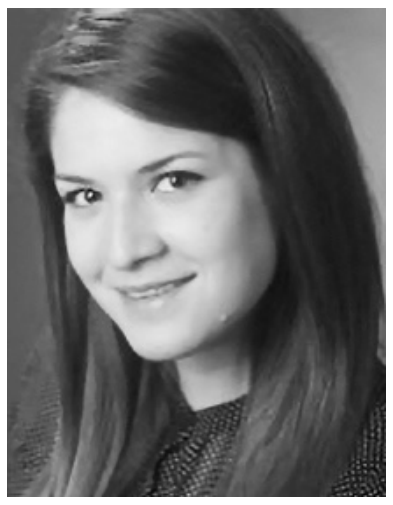

**Laura Milena Daza Parra, MSc., BSc.**

Laura Daza is an engineer, currently working for BOSCH Group Germany. In 2012, she received her BSc. degree in Electrical and Electronic Engineering from the Pontificia Universidad Javeriana Pontificia Universidad Javeriana in Bogotá, Colombia. She continued her studies at the University of Rostock, which she successfully finished in 2017 with the academic degree MSc. in Computational Engineering. In her professional career, she worked in both, industrial and academic fields, such as at the Fraunhofer IGD Rostock.

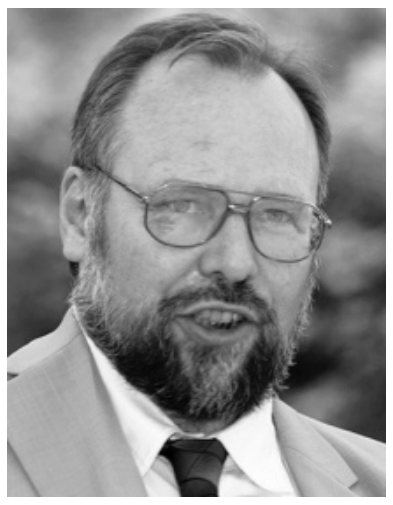

**Prof. Dr.-Ing. Bodo Urban**

Bodo Urban is currently the head of the Rostock branch of Fraunhofer IGD in Germany. He received a Diploma (Dipl.-Math.) in Mathematics and a PhD. (Dr.-Ing.) in Computer Science from the University of Rostock, Germany in 1978 and 1983, respectively. Until 1990 he was a research associate at the Computer Science Department of the University of Rostock, 1991 he moved to the new founded Rostock Division of the Computer Graphics Center, and 1992 he became responsible for the new founded Rostock Division of the Fraunhofer IGD. In 1998, he was appointed to Professor of Multimedia Communication at the Computer Science Department of the University of Rostock.